\documentclass[letterpaper]{JHEP3}
\usepackage{amsmath, amsfonts}
\usepackage{yfonts}
\usepackage{epsfig}
\usepackage{array}
\usepackage{cite}

\def\be{\begin{equation}}
\def\ee{\end{equation}}

\newcommand{\wn}{{\textswab{w}}}

\def\arccoth{\mop{arccoth}}
\def\href#1#2{#2}

\def\arctan{\mbox{arctan}}
\def\arccoth{\mbox{arccoth}}
\def\beq{\begin{equation}}
\def\eeq{\end{equation}}

\def\Ntwo{\mathcal{N}\,{=}\,2}
\def\Nfour{\mathcal{N}\,{=}\,4}
\def\alphas{\alpha_{\rm s}}
\def\Nc{N_{\rm c}}
\def\gYM{g_{\rm YM}}
\def\half{{\textstyle \frac 12}}
\def\coeff#1#2{{\textstyle \frac {#1}{#2}}}
\def\Mrest{M_{\rm rest}}
\def\Mkin{M_{\rm kin}}
\def\umin{u_{\rm min}}
\let\wn=\gamma
\def\decay{\mu}
\def\ucrit{0.02}

\preprint{{\tt hep-th/0605158}\\NSF-KITP-06-36}

\title
    {%
    Energy loss of a heavy quark moving through \boldmath $\Nfour$
    supersymmetric Yang-Mills plasma
    }

\author
    {%
    C.~P.~Herzog,\!$^1$\footnote{\email{herzog@phys.washington.edu}}~
    A.~Karch,\!$^1$\footnote{\email{karch@phys.washington.edu}}~
    P.~Kovtun,\!$^2$\footnote{\email{kovtun@kitp.ucsb.edu}}~
    C.~Kozcaz,\!$^1$\footnote{\email{kozcaz@u.washington.edu}}~
    and
    L.~G.~Yaffe$^1$\footnote{\email{yaffe@phys.washington.edu}}
    \\
    $^1$Department of Physics, University of Washington, Seattle, WA 98195--1560
    \\
    $^2$KITP, University of California, Santa Barbara, CA 93106\\
    }

\abstract{ We use the AdS/CFT correspondence to determine the rate
of energy loss of a heavy quark moving through ${\mathcal N}=4$
$SU(\Nc)$ supersymmetric Yang-Mills plasma at large 't Hooft coupling.
Using the dual description of the quark as a classical string ending
on a D7-brane, we use a complementary combination of analytic and
numerical techniques to determine the friction coefficient as a
function of quark mass. Provided strongly coupled $\Nfour$
Yang-Mills plasma is a good model for hot, strongly coupled QCD, our
results may be relevant for charm and bottom physics at RHIC. }


\begin{document}

\section{Introduction and Summary}

Energy dissipation of a heavy quark moving through a hot plasma
is both theoretically interesting and experimentally relevant
\cite{
Svetitsky:1987gq,
BraatenThoma,
Mustafa:1997pm,
Baier:2001yt,
Dokshitzer:2001zm,
Jeon:2003gi,
Mustafa:2003vh,
Armesto:2003jh,
Djordjevic:2003qk,
Djordjevic:2003zk,
MooreTeaney,
Mustafa:2004dr,
vanHees:2004gq,
Wicks:2005gt,
Adler:2005xv,
CalderondelaBarcaSanchez:2005yp,
Suaide:2005yq
}.
A high energy particle moving through a thermal medium
is an example of a non-equilibrium dissipative system.
The particle will lose energy to the surrounding medium,
leading to an effective viscous drag on the motion of the particle.
In a weakly coupled quark-gluon plasma,
the dominant energy loss mechanisms are
two-body collisions with thermal quarks and gluons,
and gluon bremsstrahlung
(see, for example, Ref.~\cite {MooreTeaney}).
Which mechanism dominates depends on the rapidity of the quark.

\subsection {Heavy ion collisions}

Most previous work on the rate of energy loss by a charged particle
moving in a plasma is based on perturbative weak-coupling approximations
\cite{Svetitsky:1987gq,BraatenThoma,MooreTeaney, 
Mustafa:2004dr,
Mustafa:1997pm,
Baier:2001yt,Jeon:2003gi,Mustafa:2003vh,Armesto:2003jh,Dokshitzer:2001zm, Djordjevic:2003zk,Djordjevic:2003qk,Wicks:2005gt},
but one would like to understand the dynamics in strongly coupled plasmas.
The specific question of the energy loss rate of a moving quark
in a strongly coupled non-Abelian plasma is of more than theoretical interest.
At RHIC, collisions of gold nuclei at 200 GeV per nucleon
are believed to produce a quark-gluon plasma which,
throughout most of the collision, should be viewed as strongly coupled
\cite{Shuryak, Shuryak:2004cy}.
For the early portion of the collision (but after apparent thermalization)
a temperature $T \approx 250$ MeV is inferred,
with a strong coupling $\alphas$ on this scale of perhaps 0.5.

Charm quarks (as observed through the production of $D$ mesons)
provide several important observables. One involves the elliptic
flow, denoted $v_2(p_t)$, which is a measure of the azimuthal
anisotropy of produced hadrons with respect to the reaction plane.
The measurement of a large elliptic flow for light hadrons is one of
the significant pieces of evidence supporting the claim that the
quark-gluon plasma produced in RHIC collisions responds like a
nearly ideal fluid with a small mean free path \cite{MolnarGyulassy,
Hirano, TeaneyLauretShuryak, TeaneyLauretShuryak2,
KolbHuovinenHeinzHeiselberg, Huovinen, MolnarHuovinen, Teaney,
Kolb:2003dz}. Since the mass of a charm quark, $m \approx 1.4$ GeV,
is large compared to the temperature, one naively expects charm
quarks to equilibrate more slowly than light quarks. Because
elliptic flow is primarily generated early in the collision, slow
thermalization of charm quarks should imply diminished elliptic flow
for charmed hadrons. The extent of the delay in thermalization, and
the resulting suppression of elliptic flow, depends on the charm
quark energy loss rate \cite{MooreTeaney}.

Another detectable effect sensitive to the rate of energy loss of quarks moving
through the plasma is jet quenching.
Within the ball of plasma formed by the collision of two large nuclei,
a quark (or antiquark) from a $q\bar q$ pair produced
near the center of the ball
is less likely to reach the edge with sufficient energy
to form a detectable jet (after hadronization) than
a quark from a $q\bar q$ pair formed near the surface of the expanding plasma.
But if one quark from a pair created near the surface escapes
and forms a jet, then the other quark, recoiling in the opposite direction,
will typically have to travel much farther through the plasma before it
can escape.
If the rate of energy loss to the plasma is sufficiently large,
then one should, and in fact does, see a suppression of back-to-back jets
(relative to $pp$ collisions).
A related quantity also sensitive to this effect
is the suppression factor $R_{AA}(p_t)$, which is the ratio
of the $D$ meson spectrum in Au-Au collisions to that in $pp$ collisions.

\subsection {$\Nfour$ supersymmetric Yang-Mills theory}

In this paper we present a calculation of the energy loss rate for
quarks moving through a plasma of
$\Nfour$ supersymmetric $SU(\Nc)$ Yang-Mills theory (SYM),
in the limit of large 't Hooft coupling, $\lambda \equiv \gYM^2 \, \Nc \gg 1$,
and a large number of colors, $\Nc \to \infty$.
The quarks whose dynamics we will study are fundamental
representation particles introduced into $\Nfour$ SYM by adding
an $\Ntwo$ hypermultiplet with arbitrary mass.%
\footnote
    {
    More explicitly, this means adding
    a Dirac fermion and 2 complex scalars,
    all in the fundamental representation,
    with a common mass and Yukawa interactions which preserve
    $\Ntwo$ supersymmetry.
    }
In the large $\Nc$ limit, fundamental representation fields have
negligible influence on bulk properties of the plasma.
One may view the quarks as test particles which serve
as probes of dynamical processes in the background $\Nfour$ plasma.

The reason for studying $\Nfour$ super-Yang-Mills is simple ---
it is easier than QCD.
There are no good approximation techniques which are generally
applicable to real-time dynamical processes in strongly coupled
quantum field theories.
Thermal relaxation or equilibration rates,
such as the energy loss rate of a moving heavy quark,
cannot be extracted directly from Euclidean correlation functions
and hence are not accessible in Monte Carlo lattice simulations.%
\footnote
    {
    See, however, Ref.~\cite{Petreczky} for a recent effort
    to extract an estimate of the damping rate
    by fitting a parametrized model of the spectral density
    to lattice data for the Euclidean current-current
    correlator.
    }
But for the specific case of $\Nfour$ $SU(\Nc)$ supersymmetric Yang-Mills
theory,
the AdS/CFT conjecture (or gauge/string duality) states
that this theory is exactly equivalent to
type IIB string theory in an $AdS_5 \times S^5$ gravitational background,
where $AdS_5$ is five dimensional anti-de Sitter space and $S^5$ is a
five dimensional sphere \cite{Maldacena, Witten, GKP}.%
\footnote
    {
    This conjecture is unproven,
    but is supported by a very large body of evidence.
    We assume its validity.
    }
At large $\Nc$ and large $\lambda$, the string
theory can be approximated by classical type IIB supergravity.
This approximation
allows completely nonperturbative calculations in the quantum
field theory to be mapped into problems in classical general relativity.
In this context, raising the temperature of the gauge theory corresponds
to introducing a black hole (or more precisely, a black brane)
into the center of $AdS_5$ \cite{Wittenblackhole}.
According to the AdS/CFT dictionary, the Hawking temperature of the
black hole becomes the temperature of the gauge theory.

At zero temperature,
the properties of $\Nfour$ supersymmetric
Yang-Mills theory are completely different from QCD.
$\Nfour$ SYM is a conformal theory with no particle spectrum
or $S$-matrix, while QCD is a confining theory with a sensible
particle interpretation.
But at non-zero temperatures
(and sufficiently high temperatures in the case of QCD),
both theories describe hot, non-Abelian plasmas with Debye screening,
finite spatial correlation lengths, and qualitatively similar
hydrodynamic behavior \cite{Kovtun:2004de}.
The major difference is that all excitations in $\Nfour$ SYM plasma
(gluons, fermions, and scalars) are in the adjoint representation,
while hot QCD plasma only has adjoint gluons and fundamental representation
quarks.
There are a variety of reasons to think that many properties of
strongly coupled non-Abelian plasmas may be insensitive to
details of the plasma composition or the precise interaction strength.
In $\Nfour$ SYM, bulk thermodynamic quantities such as the
pressure, energy or entropy densities, as well as transport coefficients
such as shear viscosity, have finite limits as the
't Hooft coupling $\lambda \to \infty$.
The pressure divided by the free Stefan-Boltzmann limit
(which effectively just counts the number of degrees of freedom)
in $\Nfour$ SYM is remarkably close to the corresponding ratio in
QCD at temperatures of a few times $T_{\rm c}$ where it is strongly coupled
\cite{GubserF}.
The dimensionless ratio of viscosity divided by entropy density
equals $1/4\pi$ in $\Nfour$ SYM, as well as in all other theories
with gravity duals \cite{kovtun, Buchel} in the strong 't Hooft coupling
limit.
And this value, which is lower than any weakly coupled theory
or known material substance \cite{Kovtun:2004de}, is in good agreement with
hydrodynamic modeling of RHIC collisions \cite {Shuryak,Shuryak:2004cy}.
These features have led some authors to speculate about
``universal'' properties of strongly coupled plasmas.

In the dual gravitational description, a fundamental representation
hypermultiplet corresponds to the addition of a D7-brane to the
black hole geometry.
This D7-brane wraps an $S^3 \subset S^5$ and wraps all of the
Schwarzschild-AdS geometry down to a minimal radius
(which is dual to the mass of the quark) \cite{kk}.
The addition
breaks the amount of supersymmetry in the theory down to ${\mathcal N}=2$.
According to the AdS/CFT dictionary, an open string whose endpoints lie
on the D7-brane is a meson, with the endpoints of the string representing
the quarks.
At non-zero temperature, one can also have open strings which stretch
from the D7-brane down to the black hole horizon.
The existence of such solutions reflects the fact that $\Nfour$
super-Yang-Mills theory, at any non-zero temperature,
is a deconfined plasma in which test quarks and antiquarks
are not bound by a confining potential.
We will extract the energy
loss rate of moving quarks in this gauge theory
by studying the behavior of the endpoints of
both types of such open string configurations.
To our knowledge, this is the first quantitative, nonperturbative
calculation of the energy loss rate of a moving massive quark
in any strongly coupled quantum field theory.\footnote{
See Ref.~\cite{SinZahed} for an interesting qualitative attempt
to understand jet quenching via AdS/CFT.
}

\subsection{A toy model, and the plan of attack}
\label{toy}

The following toy model helps to clarify some of the issues which
will arise in our analysis of heavy quark damping.
Consider a particle with momentum $p$
moving in a viscous medium and subject to a driving force $f$.
Its equation of motion may be modeled as
\be
 \dot p = -\mu \, p + f \,,
\ee
where $\mu$ is the damping rate (or friction coefficient).
To infer information about $\mu$ from motion of the particle,
it is useful to consider two different situations.
First, if one examines steady state behavior under a constant driving force,
then $\dot p = 0$ implies that $p = f/\mu$.
If the particle has an (effective) mass $m$ and its motion is
non-relativistic, so that $p = m v$, then the limiting speed
$v = f/(m \mu)$.
Hence, a measurement of the steady state speed
for a known driving force determines the combination $m \mu$,
but not $\mu$ alone.

Second, if the driving force $f=0$, then a non-zero
initial momentum will relax exponentially with a decay rate of $\mu$,
$p(t) = p(0) \, e^{-\mu t}$.
If momentum is proportional to velocity, then the speed of the particle
will show the same exponential relaxation.
A measurement of $\dot p/p$, or $\dot v/v$, will thus determine
the damping rate $\mu$.
The important point is that this second scenario is insensitive
to the value of the mass $m$.

    We will mimic these two gedanken experiments in our
analysis of open string dynamics in the $AdS_5$ black hole background.
In Section~\ref{sec:setup} we introduce notation,
describe the geometry explicitly, and derive the relevant
equations of motion for an open string.
We examine single quark solutions in Section~\ref{sec:single}.
We first find and discuss a stationary analytic solution
which may be viewed as describing a quark, of any mass,
moving in the presence of a constant external electric field
whose energy (and momentum) input precisely balances the energy
and momentum loss due to plasma damping.
Hence, this stationary solution provides
the answer to the first gedanken experiment, and yields
a measure of the (effective thermal) quark mass times the damping rate $\mu$.
We then turn to the analogue of the second gedanken experiment,
and analyze the late time behavior of a moving quark in the absence
of any external force.
Looking at the low velocity, late time behavior
allows us to linearize the string equation of motion about the
static solution, reducing
the problem to a calculation of the quasi-normal modes of the resulting
linear
operator.
This second gedanken experiment yields
the damping rate $\mu$ directly.

Both the stationary analytic solution, and the quasi-normal mode
analysis involve strings which are sensitive to the geometry
arbitrarily close to the black hole horizon.
This near horizon dependence
turns up a number of subtle issues involving
infrared sensitivity and the extent to which the total energy
of a moving quark is well-defined.
These issues are discussed in Section~\ref{sec:single},
but to insure that our interpretation is sensible,
in Section~\ref{sec:numerics} we study quark-antiquark solutions,
or string configurations in which both endpoints lie on the D7-brane.
These mesonic configurations
avoid the infrared subtleties of the single quark solutions,
but at the cost of requiring the numerical solution of nonlinear
partial differential equations.
Nevertheless, we are able to find back-to-back quark/antiquark
solutions with external forcing in which the quark and antiquark move
apart at constant velocity,
as well as unforced back-to-back solutions in which the quark
and antiquark decelerate while separating monotonically.
The damping rate may be extracted from these
quark/antiquark solutions when the particles are widely separated,
and the results confirm and extend the previous analysis based
on single quark solutions.

We discuss some conceptual issues, including the effects of
fluctuations which must inevitably accompany dissipation
in Section~\ref{sec:discussion}.
A number of other interesting analytic solutions are briefly described
in Appendix~\ref{app:other}.
Appendix~\ref{app:qnm} presents the result of the quasi-normal mode
analysis in three dimensions, where a completely analytic treatment is possible.
The final Appendix~\ref{app:error} briefly discusses integration technique
and numerical error in the non-linear PDE solutions of
Section~\ref{sec:numerics}.

\subsection{Summary of results}

For a quark moving with arbitrary velocity $v$,
we find that the rate at which it loses energy and momentum to the plasma
is given by
\be
\label{eq:EPrate}
    \frac{dp}{dt}
    =
    \frac 1v \, \frac{dE}{dt}
    =
    -
     \frac \pi 2 \, \sqrt\lambda \, T^2 \,
    \frac{v}{\sqrt{1-v^2}} \,.
\ee
This momentum loss rate (or viscous drag), as a function of velocity,
is independent of the quark mass.
Note that the viscous drag
may also be interpreted as
the energy loss per unit distance traveled, since
$
    \frac{dE}{dx} = \frac 1v \, \frac{dE}{dt}
$.

Re-expressing the viscous drag in terms of momentum, instead of velocity,
requires knowledge of the dispersion relation relating the
energy $E$ and momentum $p$, and hence the relation between
velocity $v = dE/dp$ and momentum.
As discussed below,
the scale of thermal corrections to the energy of the quark
in the strongly-coupled $\Nfour$ medium is given by
\be
    \Delta m(T) \equiv \half \sqrt\lambda T \,.
\ee
A heavy quark should be viewed as one whose mass $m$ is large compared to
$\Delta m(T)$ (not just large compared to $T$).%
\footnote
    {
    In theories with unbroken $\Ntwo$ supersymmetry the Lagrangian
    (or bare) mass does not get renormalized.
    So there is no need to distinguish between bare and
    renormalized mass, or deal with scale dependence in the mass ---
    it is a physical parameter.
    The heavy mass regime, $m \gg \Delta m(T)$,
    may equivalently be viewed as the low temperature regime,
    $T \ll 2m/\sqrt\lambda$.
    In this form, one sees that the relevant scale
    which distinguishes low versus high temperature
    is not $m$, but rather $m/\sqrt\lambda$.
    This is the scale of the mass of deeply bound $q\bar q$ states which
    form in zero temperature $\Nfour$ SYM with massive fundamental
    representation quarks \cite{myers}.
    So low temperature corresponds to the regime where these
    mesonic bound states form a dilute, non-relativistic gas.
    }
If $m \gg \Delta m(T)$, then thermal corrections to the zero-temperature
relativistic dispersion relation are negligible.
In this regime, the viscous drag (\ref {eq:EPrate}) is equivalent to
\be
    \frac{dp}{dt} = -\mu \, p
    \label{ploss}
\ee
with a friction coefficient
\begin{equation}
    \mu = \pi T \> \frac {\Delta m(T)}{m}
    \qquad
    \mbox{[heavy quarks, $m \gg \Delta m(T)$]} \,. \hspace*{-2cm}
\end{equation}
The momentum independence of the friction coefficient is a surprising
result which differs from the behavior of a weakly coupled plasma.%
\footnote
    {
    For a heavy quark moving through a weakly coupled plasma,
    the dominant mechanism of energy loss is two body scattering
    off thermal quarks and gluons,
    provided $1-v^2$ is not parametrically small.
    The resulting loss rate $\mu$ is a non-trivial function
    of velocity \cite{MooreTeaney}.
    Only for small velocity is the energy loss rate
    well modeled by Eq.~(\ref{ploss}) with a constant value of $\mu$.
    For ultra-relativistic quarks
    the dominant scattering process switches from two-to-two
    scattering events, in which the momentum transfer is a small
    fraction of the heavy quark momentum, to gluon bremsstrahlung
    in which each gluon emission can change the heavy quark momentum
    by an $\mathcal O(1)$ amount.
    In this regime, characterizing the energy loss as a smooth
    differential process, as in Eq.~(\ref{ploss}), no longer
    makes sense.
    }

The dispersion relation of lighter quarks, for which the ratio
$m/\Delta m(T)$ is of order one, is substantially influenced by the
medium. A quark at rest in the medium corresponds to a straight
string stretching from the D7-brane down to the horizon. Such a
quark, immersed in the thermal $\Nfour$ medium at temperature $T$,
has a rest energy $\Mrest(T)$ which differs from its Lagrangian mass
$m$. Determining this relation requires solving (numerically) for
the change in the embedding of the D7-brane induced by the black
hole horizon \cite{Babington:2003vm,andy}.
Asymptotically,%
\footnote
    {
    The coefficient of the $(\Delta m/m)^{8}$ term in Eq.~(\ref{eq:Mrest})
    was determined by a numerical fit, not analytically,
    and may not be exact.
    }
\begin{equation}
    \Mrest(T)
    = m \left\{ 1 - \frac {\Delta m(T)}{m}
    +\frac{1}{8} \Big(\frac {\Delta m(T)}{m}\Big)^{\!4}
    -\frac{5}{128}  \Big(\frac {\Delta m(T)}{m}\Big)^{\!8}
    + {\mathcal O}\!\left[\Big(\frac {\Delta m(T)}{m}\Big)^{\!12}\right]
    \right\}
    \,.
\label{eq:Mrest}
\end{equation}
Truncating after three terms gives a result which is
accurate for $\Mrest+\Delta m$ 
to within 1\%.

As $m/\Delta
m(T)$ approaches a critical value of approximately 0.92,
the thermal rest mass nearly vanishes.%
\footnote
    {
    The minimal value of $\Mrest$ at this point is $\ucrit \Delta m(T)$.
    When one decreases the mass beyond this point
    the location of the D7-brane
    jumps discontinuously to the horizon and
    $\Mrest$ vanishes.
    See Refs.~\cite{Babington:2003vm,andy,Mateos:2006nu,Albash:2006ew}
    for discussion of this transition.
    }
Our semiclassical string analysis is only valid when the zero
temperature mass exceeds this critical value.
The resulting dependence is plotted in Figure~\ref{fig:masses}.

\begin{FIGURE}[t]
{
   \centerline{\psfig{figure=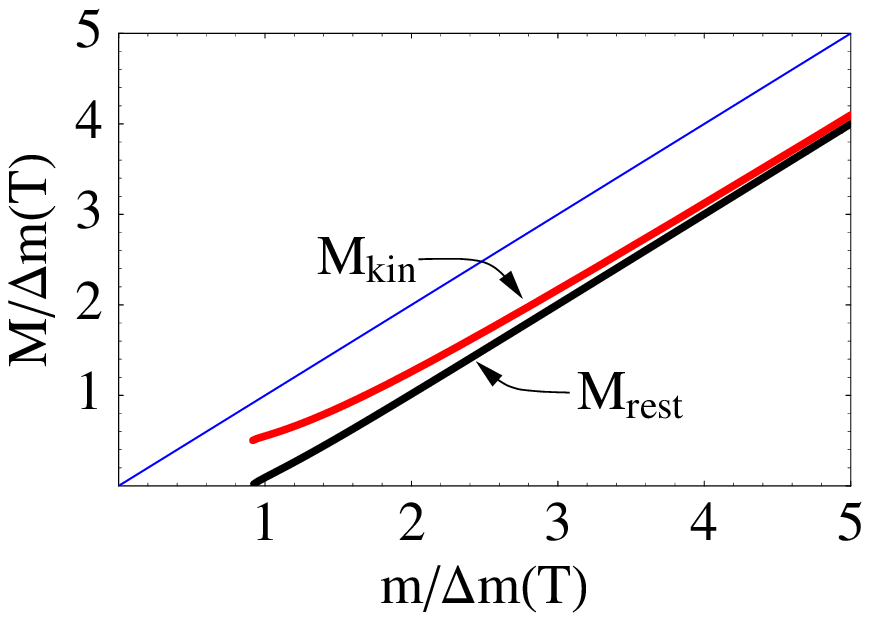,width=3.5in}}
   \vspace*{-20pt}
    \caption{The thermal rest mass (or energy) $\Mrest$
    and the kinetic mass $\Mkin$
    of a quark immersed in the $\Nfour$ plasma at temperature $T$,
    as functions of the zero-temperature mass $m$,
    with all masses measured in units
    of $\Delta m(T) = \half\sqrt\lambda T$.
    At $m \approx 0.92 \, \Delta m(T)$, the location of the D7-brane 
    jumps discontinuously to the horizon
    \cite{Babington:2003vm,andy,Mateos:2006nu,Albash:2006ew}.
    }
\label{fig:masses}
}
\end{FIGURE}

The dispersion relation of a moving quark in
the thermal medium need not be Lorentz invariant
since the plasma defines a preferred rest frame.
For non-relativistic motion, the dispersion relation will have the form
\be
    E(p) = \Mrest(T) + \frac {p^2}{2\Mkin(T)} + \mathcal O(p^4) \,.
\label{eq:NRdispersion}
\ee
The effective kinetic mass $\Mkin(T)$ is not the same
as the thermal rest mass $\Mrest(T)$.
For heavy quarks, we find that $\Mkin(T)$ differs negligibly
from the thermal rest mass,
\be
    \Mkin(T) = \Mrest(T)
    + {\mathcal O}\Big[ m \Big(\frac {\Delta m(T)}{m}\Big)^{\!2} \, \Big]\,,
\ee
but the difference between these masses becomes significant as
the quark mass decreases.
The dependence of the kinetic mass $\Mkin(T)$ on the quark mass
is also shown in Figure~\ref{fig:masses}.
As $m/\Delta m(T)$ approaches the lower critical value of 0.92,
the kinetic mass $\Mkin(T)$ has
a limiting value just slightly greater than $\half \Delta m(T)$.
As $m/\Delta m(T) \to \infty$, both $\Mkin$ and
$\Mrest$ approach $m - \Delta m(T)$.

For not-so-heavy quarks moving relativistically,
we can only infer the dispersion relation from analysis
of the time-dependent numerical solutions discussed in
Section~\ref{sec:numerics}.
We do not have any analytic derivation, but all our numerical
results are consistent with the thermal dispersion relation
\be
    E(p) = \Mrest(T) - \Mkin(T) + \sqrt{p^2 + \Mkin(T)^2} \,,
\label{eq:dispersion}
\ee
which reduces to Eq.~(\ref{eq:NRdispersion}) for low momentum,
and gives $v \equiv dE/dp = p / \sqrt{p^2 + \Mkin(T)^2}$.
For this relation between velocity and momentum,
the viscous drag (\ref {eq:EPrate}) is equivalent to
$\dot p = -\mu p$ with a friction coefficient
\be
    \mu
    = \pi T \> \frac {\Delta m(T) }{\Mkin(T)}
    = \frac \pi 2 \, \frac{\sqrt{\lambda} \, T^2}{\Mkin(T)}
    \,.
\label{result}
\ee
As the quark mass decreases, the
kinetic mass $\Mkin(T)$ has a lower limit of $\half \Delta m(T)$,
and hence the friction coefficient has a remarkably simple
upper limit,%
\footnote
    {
    Strictly speaking,
    $\mu$ equals $2\pi T$ only in the limit where $\Mrest=0$,
    {\em i.e.}, for an unstable D7-brane configuration.
    As the D7-brane is stable already for
    $\Mrest=0.02 \Delta m(T)$,
    this limiting value of $\mu$ is very close to the actual value
    for the lightest accessible quark masses.
    }
\begin{equation}
    \mu \le 2\pi T \,,
\end{equation}
which turns out to be dimension independent.
It is tempting to speculate,
along the lines of Ref.~\cite{Kovtun:2004de},
that the ratio $\mu / T$ is bounded above by $2 \pi$
even in more general theories.

Knowledge of the viscous drag on a quark is equivalent to knowledge
of the diffusion constant for quark ``flavor''.
The relation is $D = T/(\mu \Mkin)$,
so our result (\ref{result}) is equivalent to a flavor diffusion
constant
\begin {equation}
    D =  \frac 1{\pi \Delta m(T)}
    = \frac 2 {\pi \sqrt\lambda \, T} \,.
 \label{diffusionresult}
\end {equation}
As discussed in Section \ref{sec:discussion},
this same information may also be recast
as the rate of change of the mean square transverse momentum
of a quark initially moving in a given direction,%
\footnote
    {
    This result for the rate of change of mean square transverse
    momentum follows from modeling the effects of fluctuations
    in the momentum of the quark by a simple Langevin equation
    in which the noise, characterizing stochastic fluctuations
    in the force exerted on the quark,
    has an isotropic velocity-independent variance.
    At weak coupling \cite {MooreTeaney},
    the friction coefficient and
    the stochastic force variance both show significant velocity
    dependence unless $v \ll 1$.
    Our strong coupling result (\ref{result}) for the friction coefficient is
    velocity independent and valid for arbitrary values of the
    quark's rapidity,
    but we do not have a direct determination
    of the stochastic force variance for arbitrary velocities.
    If the force variance shows the same velocity independence
    as the friction coefficient, then the result (\ref{jetquenching})
    will also be valid for arbitrary rapidity,
    but if this is not true, then the result (\ref{jetquenching})
    will only be valid for non-relativistic motion with $v \ll 1$.
    }
\begin {equation}
    \frac {d}{dt} \, \left\langle (\vec p_\perp)^2 \right\rangle
    =
    \frac {4 \, T^2}{D}
    =
    4 \pi \Delta m(T) T^2
    =
    2\pi \sqrt\lambda \, T^3 \,.
    \label{jetquenching}
\end {equation}
This quantity, divided by the velocity of the quark
(to give a rate of change per unit distance traveled)
is sometimes called the ``jet quenching parameter''
$\hat q$ \cite{Baier:1996sk}.

Physically characterizing the mechanism responsible for the energy loss
(\ref{result}) in terms of some microscopic picture of the dynamics
of the $\Nfour$ SYM field theory is a challenge.
In the AdS dual description, energy and momentum flow along the
string which hangs down from the quark, away from the D7-brane
and toward the black hole horizon.
It is clear that the portion of the string which lies close to the
horizon should be thought of as describing long distance deformations
of the medium surrounding the quark.
The energy loss should not be regarded as resulting from scattering off
excitations in the thermal medium.  Any scattering
would correspond to
small fluctuations in the string worldsheet,
and these fluctuations are suppressed by inverse powers of the
't Hooft coupling (or string tension).
Nor is the energy loss due to radiation of glueballs,
which correspond to closed strings breaking off of the open string
and whose emission is suppressed by powers of $1/\Nc$.
 For relativistic velocities, $v \to 1$,
there is nothing in the classical string dynamics which is reminiscent
of near-collinear gluon bremsstrahlung which (at weak coupling)
can cause a fast moving quark to lose an $\mathcal O(1)$ fraction
of its momentum in a single scattering.
Rather, the energy transfer from the moving quark
to the surrounding plasma should be regarded as analogous to
the formation of a wake in the coherent polarization
cloud surrounding a charged particle moving through a
polarizable medium, or the wake on the surface of water
behind a moving boat.

Ultimately, the energy transfered to the medium from the moving quark
must appear as heating and outward hydrodynamic flow in the
non-Abelian plasma surrounding the quark.
To see this flow directly, one would like to evaluate the expectation
value of $T^{\mu\nu}(x)$ in the presence of the moving quark.
The AdS/CFT correspondence provides a recipe for this calculation,
but its implementation is difficult.
The expectation value of $T^{\mu\nu}$ corresponds to
boundary fluctuations in the gravitational metric.  Deriving
these fluctuations from the energy distribution of the string requires
graviton propagators in the black hole geometry, for which
no closed form analytic expression is currently available.
Performing the computations required to evaluate
$\langle T^{\mu\nu}(x) \rangle$
would allow one to examine the form of the wake behind a moving quark and,
for example,
see the sonic boom produced by a quark moving faster
than the speed of sound.
Sadly, we leave such a study for future work.

\section{Open string dynamics in the black brane background}\label{sec:setup}

The AdS/CFT correspondence \cite{Maldacena, Witten, GKP}
posits an equivalence
between $\Nfour$ $SU(\Nc)$ supersymmetric Yang-Mills theory and type IIB string
theory in a $AdS_5 \times S^5$ background.
Type IIB strings are
characterized by two numbers: a string coupling $g_s$
and a tension $T_0$,  
or equivalently a fundamental string length scale
$\ell_s \equiv (2\pi T_0)^{-1/2}$.
The background is characterized by the radius of curvature of the $AdS_5$
and of the $S^5$,
which must be equal and will be denoted by $L$.

The AdS/CFT correspondence provides a dictionary between these two seemingly
very different physical theories.
One important entry in this dictionary is the
relationship between the string coupling and the Yang-Mills coupling,
\be
4 \pi g_s = g_{YM}^2 \ ,
\ee
and an equally important entry is the relationship between the string tension,
the radius of curvature $L$, and the 't Hooft coupling,
\be
\left({L}/{\ell_s}\right)^4 =  \lambda \equiv g_{YM}^2 \, \Nc \ .
\ee
For the convenience of readers
who are not completely conversant with the AdS/CFT correspondence,
these relations, plus additional ones which will appear as we progress,
are summarized in Table~\ref{tab:dict}.
\begin{TABLE}[t]
{
\setlength{\extrarowheight}{2pt}
\begin{tabular}{ccl}
    AdS & $\mathcal N{=}4$ SYM
    & quantity
\\\hline
    $L$ & --
    & $AdS_5$ and $S^5$ curvature radius
\\
    $\ell_s$ & $\lambda^{-1/4} L$
    & fundamental string scale [${}\equiv \sqrt{\alpha'}$]
\\
    $(L/\ell_s)^4$ & $\lambda$
    & 't Hooft coupling [${} \equiv \gYM^2 \, \Nc$]
\\
    $T_0$ & $\frac {\sqrt\lambda}{2\pi} \, L^{-2}$
    & string tension [${}\equiv (2\pi \ell_s^2)^{-1}$]
\\
    $g_s$ & $\frac 1{4\pi} \, \gYM^2$
    & string coupling
\\
    $u_h$ & $\pi T$
    & black hole horizon radius (${}\times L^{-2}$)
\\
    $u_h/\pi$ & $T$
    & temperature
\\
    $u_m$ & $\frac {2\pi}{\sqrt\lambda} \, (\Mrest{+}\Delta m)$
    & minimal radius of D7-brane (${}\times L^{-2}$)
\\
    $T_0 L^2 \, u_h$ & $\Delta m(T)$
    & thermal rest mass shift [${}= \half\sqrt\lambda T$]
\\
    $T_0 L^2 \, (u_m{-}u_h)$ & $\Mrest(T)$
    & static thermal quark mass
\end{tabular}
\caption
    {
    AdS/CFT translation table.
    The static thermal quark mass $\Mrest(T)$ is the free energy
    of quark at rest in the $\Nfour$ SYM plasma.
    It equals the Lagrangian quark mass $m$ in the zero temperature limit.
    }
\label{tab:dict}
}
\end{TABLE}

\subsection{Adding a black hole}

The gravity dual of finite temperature ${\Nfour}$ $SU(\Nc)$ super
Yang-Mills theory is $S^5$ times the five dimensional AdS-black hole
solution \cite{Wittenblackhole}.
This solution is a geometry in which a black hole (or more properly,
a black brane with a flat four dimensional horizon) is placed
inside AdS space.
The metric of the resulting AdS black brane solution
in $d+1$ dimensions may be written as
\beq
\label{stmetric}
ds^2 =  L^2 \left (\frac{du^2}{h(u)} - h(u)\, dt^2 + u^2 \delta_{ij}\, dx^i dx^j
 \right ) ,
\eeq
where
\beq
    h(u) = u^2 \left[1 - \left(\frac{u_h}{u} \right)^d \right] .
\eeq
Since the case of arbitrary dimension is usually as easy
to compute as the specific $d=4$ case of interest,
we will leave $d$ arbitrary in much of this section.
Our radial coordinate $u$ has been rescaled by a factor of $L^{-2}$;
some authors use $r = L^2 \, u$ instead \cite{magoo}.
The black hole horizon is located at $u=u_h$ where $h(u)$ vanishes.

The Hawking temperature of the black hole
equals the temperature of the field theory dual.
The horizon  radius is related to the Hawking temperature by
\beq T =  \frac{d}{4 \pi } \, u_h \ ,
\label{TH}
\eeq
or $u_h = \pi \, T$ in $d=4$.

Introducing a flavor of fundamental representation quarks corresponds,
in the gravity dual of the four dimensional field theory,
to the addition of a D7-brane \cite{kk}.  This D7-brane
wraps an $S^3$ inside the transverse $S^5$ and fills all of the asymptotically
$AdS$ space down to some minimum radial value $u = u_m$.
We require that $u_m > u_h$.
Given an open string that ends on this D7-brane,
the quark is reinterpreted as the string's endpoint.
The choice of $u_m$ is equivalent to the choice of mass for the quark;
the relation between $u_m$ and quark mass will be discussed in
Section~\ref{sec:single}.

\subsection{String equations of motion}

The dynamics of an open string ending on the D7-brane
depends on the background geometry, but the back reaction
of the string on the geometry is negligible and may be neglected.
The negligible back reaction
reflects the fact that fundamental representation
quarks only make an $\mathcal O(\Nc)$ contribution to the free energy
which is small, in the $\Nc\to\infty$ limit,
relative to the $\mathcal O(\Nc^2)$
contributions of the adjoint representation fields of $\Nfour$ SYM.

The dynamics of a classical string is governed by the Nambu-Goto action,
\beq
S = -T_0 \int d \sigma \> d\tau \sqrt{-\operatorname{det}{g_{ab}}} \ .
\eeq
The coordinates $(\sigma, \tau)$ parametrize
the induced metric $g_{ab}$ on the string
world-sheet.
Let $X^\mu(\sigma, \tau)$ be a map from the string world-sheet into space-time,
and define
$\dot X = \partial_\tau X$, $X' = \partial_\sigma X$, and $V \cdot W = V^\mu W^\nu G_{\mu\nu}$
where $G_{\mu\nu}$ is the space-time metric (\ref{stmetric}).  Then, writing
$\operatorname{det}{g_{ab}} = g$, one has
\be
-g = (\dot X \cdot X')^2 - (X')^2(\dot X)^2 \ .
\ee
We will limit our attention to strings which lie within a three dimensional
slice of the asymptotically AdS space in which all but one (call it $x$)
of the transverse coordinates $x^i$ vanish.
So $X(\sigma,\tau)$ will be a map to $(t, u, x)$.
Choosing a static gauge where $\sigma = u$ and $\tau=t$,
the string worldsheet is described by a single function $x(u,t)$.
With this choice, one finds that
\begin{subequations}
\begin{eqnarray}
\dot{X} \cdot X' &=& L^2 \left( u^2 \dot{x} x'  \right) , \\
 (X')^2 &=& L^2 \left[ h^{-1} + u^2 (x')^2 \right] , \\
 (\dot{X})^2 &=& L^2 \left[ - h + u^2 (\dot{x})^2  \right] ,
\end{eqnarray}
\end{subequations}
and the induced metric becomes
\beq g_{ab} = L^2 \left [ \begin{array}{cc}
         - h + u^2(\dot{x})^2 & u^2 \, \dot{x} x' \\ u^2 \, \dot{x} x' &
         \frac{1}{h} + u^2 (x')^2
         \end{array} \right ] \ .
\eeq
The determinant of $g_{ab}$ is
\beq
-\frac{g}{L^4} = 1 - h^{-1} \, {u^2} (\dot{x})^2 + h\, u^2 (x')^2 \ .
\label{eq:-g}
\eeq
 From this determinant, the
equation of motion that follows from the Nambu-Goto action is
\beq
\label{eomt}
\frac{\partial}{\partial u} \left( h \, u^2 \frac{x'}{\sqrt{-g}} \right)   -
\frac{u^2}{h} \frac{\partial}{\partial t}
\left( \frac{\dot{x}}{\sqrt{-g}} \right) =0 \ .
\eeq

Useful to us in the following are general expressions for the canonical momentum
densities associated to the string,
\begin{subequations}
\begin{eqnarray} \label{raw0current}
\pi^0_{\mu} &=& - T_0 \, G_{\mu \nu} \,\frac{(\dot{X} \cdot X') (X^{\nu})'
-(X')^2 (\dot{X}^{\nu})}{\sqrt{-g}} \ , \\
\pi^1_{\mu} &=&  -T_0 \, G_{\mu \nu} \,\frac{(\dot{X} \cdot X') (\dot{X}^{\nu})
-(\dot{X})^2 (X^{\nu})'}{\sqrt{-g}} \ .
\end{eqnarray}
\end{subequations}
For our string, these expressions reduce to
\begin{equation}
\label{densities}
    \left(
    \begin{array}{c}
    \pi^0_x \\ \pi^0_u \\ \pi^0_t
    \end{array}
    \right)
    =
    \frac{T_0 L^4}{\sqrt{-g}}
    \left[
    \begin{array}{c}
    \dot x \, u^2 \, h^{-1} \\
    -\dot x \, x' \, u^2 \, h^{-1} \\
    -1 - (x')^2 \, u^2 \, h
    \end{array}
    \right] ,
    \quad
    \left(
    \begin{array}{c}
    \pi^1_x \\ \pi^1_u \\ \pi^1_t
    \end{array}
    \right)
    =
    \frac{T_0 L^4}{\sqrt{-g}}
    \left[
    \begin{array}{c}
    -x' \, u^2 \, h \\
    -1 + (\dot x)^2 u^2 \, h^{-1} \\
    \dot x \, x' \, u^2 \, h
    \end{array}
    \right] .
\end{equation}
The density of energy and $x$-component of momentum on the string
worldsheet are given by $\pi^0_{t}$ and $\pi^0_{x}$, respectively.
Integrating them along the
string gives the total energy and momentum of the string,
\be
E = -\int d\sigma \> \pi_t^0 \,,\qquad p = \int d\sigma \>  \pi_x^0 \ .
\label{EPeqs}
\ee

\section{Single quark solutions}
\label{sec:single}

\subsection {Static strings}
\label{sec:static}

Single quark solutions correspond to strings which hang from the
D7-brane down to the black hole horizon.
The simplest solution to the string equation of motion (\ref {eomt})
is just a constant, $x(u,t) = x_0$.
This solution describes
a static string stretching from $u = u_m$ straight down
to the black hole horizon at $u_h$,
and clearly represents a static quark at rest in the thermal medium.
We may compute the energy and momentum
of such a configuration using Eq.~(\ref{EPeqs}).
The energy
\be
E = T_0 L^2 \int_{u_h}^{u_m} du  = T_0 L^2 \, (u_m-u_h) \,,
\label{eq:Estatic}
\ee
while the total momentum $p$ (and momentum density $\pi^0_x$) vanish.
This energy must equal the (Lagrangian) mass $m$ of the quark
in the zero temperature limit.
Recalling, from Eq.~(\ref{TH}), that $u_h$ is proportional to the temperature,
we see that
\be
    T_0 L^2 \, u_m = m \qquad \mbox{(zero temperature)} \,.
\label{eq:zero T mass}
\ee
Moving the D7-brane to a larger radius (larger $u_m$)
increases the mass of the quark;
a D7-brane sitting at the boundary of the (asymptotically) AdS space
corresponds to quarks of infinite mass.

However, raising the temperature affects the relation
between the Lagrangian mass $m$ and the position $u_m$ of the D7-brane
in the gravitational description.%
\footnote
    {
    The D7-brane is a dynamical object whose equations of motion
    are equivalent to minimizing its worldvolume.
    The tension of the brane is a negligible perturbation to the
    background black hole geometry (suppressed by $1/\Nc$),
    but the D-brane does respond to variations in the background geometry
    by changing its embedding.
    According to the AdS/CFT dictionary the mass associated to
    a given embedding is determined by the asymptotic form of the
    D7-brane configuration. For zero temperature this
    procedure reproduces Eq.~(\ref{eq:zero T mass}).
    }
The result has the form
\begin{equation}
    \frac{T_0 L^2 \, u_m}{m}
    =
    1 + g\Big(\frac{T_0 L^2 u_h}{m}\Big) \,,
\label{eq:um/m}
\end{equation}
with the correction $g(x)$ behaving (for $d=4$) as
\begin{equation}
    g(x) = \frac{1}{8} \, x^4 - \frac{5}{128} \, x^8 + \mathcal O(x^{12})\,.\end{equation}
Retaining just the first two terms in $g(x)$
gives a result for $T_0 L^2 u_m/m$
which is accurate to within 1\%.

The energy (\ref{eq:Estatic}) of the static string
should be interpreted as the free energy
of a static quark sitting in the thermal $\Nfour$ SYM medium.%
\footnote
    {
    This static string configuration describes the expectation value
    of a fundamental representation Wilson line in thermal $\Nfour$ SYM
    \cite{brandhuber,reyt}.
    In the finite temperature field theory, this expectation value gives
    $e^{-\beta F_q}$, where $F_q$ is the free energy, not the internal energy,
    of a static quark.
    }
In a small abuse of language,
we will refer to this free energy as the static thermal mass, $\Mrest(T)$.
Using $u_h = \pi T$ [from Eq.~(\ref{TH})]
plus $T_0 L^2 = \frac{\sqrt{\lambda}}{2\pi}$,
the result is
\begin{subequations}
\label{Ezero}
\begin{eqnarray}
    \Mrest(T) &\equiv&
    T_0 L^2 \, (u_m-u_h)
\nonumber
\\ &=&
    m - \Delta m(T) + m \, g\Big(\frac{\Delta m(T)}{m}\Big) \,,
\\
\noalign {\hbox{with}}
    \Delta m(T) &\equiv& T_0 L^2 u_h = \half \sqrt{\lambda} T \,.
\end{eqnarray}
\end{subequations}
This relation between the static thermal mass $\Mrest(T)$
and the Lagrangian mass $m$ is plotted in Figure~\ref{fig:masses}.

\subsection {Moving, straight strings}

A rigidly moving string profile $x(u,t) = x_0 + vt$ is also a solution
to the string equation of motion (\ref {eomt}).
However, such rigid motion of the string is not physical.
The problem is that $-g$ is not positive definite for this profile.
One finds that $g$ vanishes at a critical value $u_c$ given by
\beq
\label{critical}
(u_c)^{d} = \frac{(u_h)^{d}}{1-v^2} \,.
\eeq
For any non-zero velocity, $u_c > u_h$
and $-g$ is negative in the region $u_h < u < u_c$ between the
horizon and the critical value of the radius.
A negative determinant is often a signal of superluminal propagation.
When $g = 0$ the induced metric on the string world sheet is degenerate, and
if $-g<0$ then the action, energy, and momentum all become complex,
which means this solution must be discarded.
By choosing
$x=vt$, we picked inconsistent initial conditions where
parts of the string have a velocity faster than the local speed of light
at $t=0$. While time evolution of this initial configuration gives
a very simple solution, it is not physical.

\subsection{Moving, curved strings}
\label{sec:analytic}

To find a physical configuration which corresponds to a quark moving
at constant velocity,
we will look for stationary solutions of the form
\beq
x(u,t) = x(u) + v t \,.
\label{analconstantt}
\eeq
For string profiles of this form,
$x'$, $\dot{x}$, and $\sqrt{-g}$ are time-independent.
The time derivative term in the equation of motion (\ref{eomt})
completely drops out and we are left with an ordinary differential equation
\beq
\label{ourdiffeq}
 \frac{d}{du} \left ( h \, u^2 \frac{x'}{\sqrt{-g}}
\right )   =0
\eeq
where
\beq
\sqrt{-g}=L^2 \left[ 1 - h^{-1} \, u^2 \, v^2 + h\, u^2\, (x')^2 \right]^{1/2}
\,.
\eeq
This differential equation is straightforward to integrate.
(The $v=0$ limit of this equation
appeared in the finite temperature calculation
\cite{brandhuber,reyt} of the AdS Wilson line
\cite{mwilson,rwilson}.)
The first integral is
\beq
    \frac{x'}{\sqrt{-g/L^4}} \, = \frac{C \, v}{u^2 \, h} \,,
\eeq
or
\begin{equation}
\label{oursol}
    (x')^2
    =
    \frac {C^2 \, v^2} {u^{8}} \left[ 1- (u_h/u)^d \right]^{-2} \>
    \frac {1 -v^2- (u_h/u)^d }{1- C^2 \, v^2 \, u^{-4}-(u_h/u)^d }
    \,,
\end{equation}
where $C$
is an integration constant.  This constant determines
the momentum current flowing along the string.
From the expressions (\ref{densities}) for these currents, we see that
\be
\pi_x^1 = -T_0 L^2 \, C \, v \,, \qquad \pi_t^1 = T_0 L^2 \, C \, v^2 \,,
\label{eq:pi}
\ee
showing that these two currents are constant along the length of the string.

Solving for $-g$ yields
\beq
\frac{-g}{L^4} = \frac{ 1- v^2 - (u_h/u)^d }{1
-C^2 \, v^2 \, u^{-4}-(u_h/u)^d } \ .
\eeq
Both numerator and
denominator are positive for large $u$, and negative for $u$ near $u_h$.
So the only way for $-g$ to remain positive everywhere on
a string that stretches from the boundary to the horizon
is to have both numerator and denominator change sign at the same point.%
\footnote
    {
    There are other solutions to these equations which
    have a turn-around point and do not correspond to strings
    running from large $u$ down to the horizon.
    We discuss these other solutions in Appendix~\ref{app:other}.
    }
This condition
uniquely fixes $C$ up to a sign,
\beq
C = \pm\left ( \frac{u_h^d}{1-v^2} \right )^{2/d}
\quad \mathop{\longrightarrow}\limits^{d=4} \quad
\pm\frac{u_h^2}{\sqrt {1-v^2}} \ .
\eeq
For the specific case of $d=4$,
$-g/L^4$ reduces to $1-v^2$, and we have
\beq
\label{derrel} x'(u) = \pm v \, \frac{u_h^2}{h(u) \, u^2}\ .
\eeq
Integrating $x'$ yields solutions of the form
\begin{equation}
    x(u,t) = x_\pm(u,t) \equiv x_0 \pm v \, F(u;v^2) + vt \,,
\label{eq:analytic}
\end{equation}
where the upper sign corresponds to $C$ positive; the lower to $C$ negative,
and $x_0$ is the position of the string at the $u=\infty$ boundary at
time zero.
For arbitrary $d$, $F(u;v^2)$ is a hypergeometric function.
For $d=4$, this function reduces to the velocity independent expression,
\beq
F(u) =\frac{1}{2 \, u_h} \left [\frac {\pi}2 -
\arctan \!\left(\frac{u}{u_h} \right) -
\arccoth \!\left(\frac{u}{u_h} \right) \right ] .
\label{analyticsol}
\eeq
This function is plotted in Figure~\ref{heavystring}.
It vanishes as $u\to\infty$ and diverges to $-\infty$ as $u \to u_h$;
its asymptotic behavior is
\begin{equation}
    F(u) = \begin{cases}
    \displaystyle
    -\frac {u_h^2}{3u^3} + O\Big(\frac{u_h^6}{u^7} \Big) \,,
    & u \to \infty \,;
    \\[10pt]
    \displaystyle
    -\frac 1{4u_h} \ln \Big(\frac {2u_h}{u{-}u_h}\Big)
    + \frac \pi {8u_h}
    + O\Big(\frac{u{-}u_h}{u_h^2} \Big) \,,
    & u \to u_h \,.
    \end{cases}
\end{equation}

\begin{FIGURE}[t]
{
   \centerline{\psfig{figure=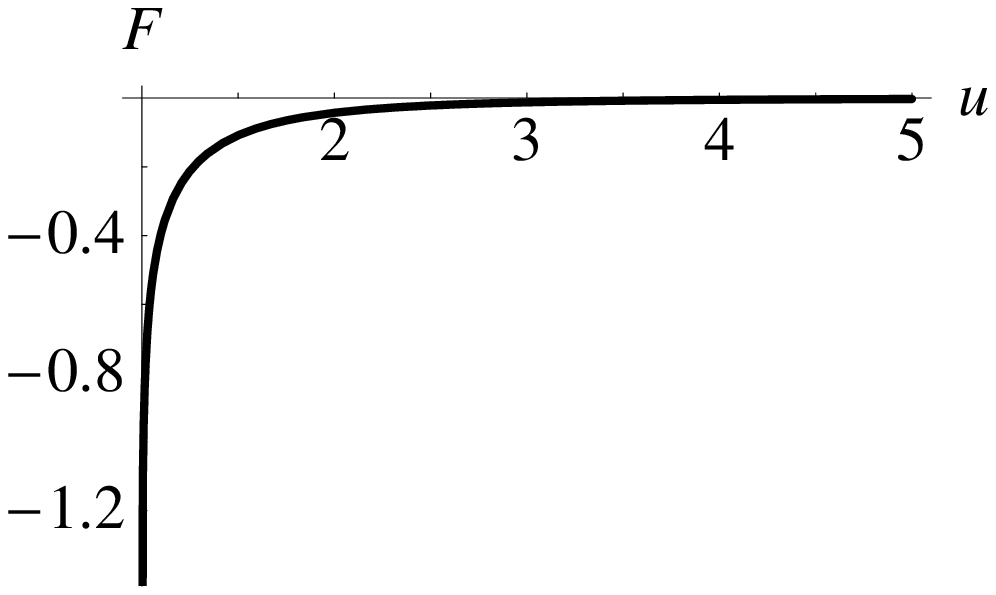,width=1.8in}}
    \caption{A plot of the function $F(u)$ which determines the
    string profile, in units where $u_h = 1$.}
\label{heavystring}
}
\end{FIGURE}

\begin{FIGURE}[t]
{
   \centerline{\psfig{figure=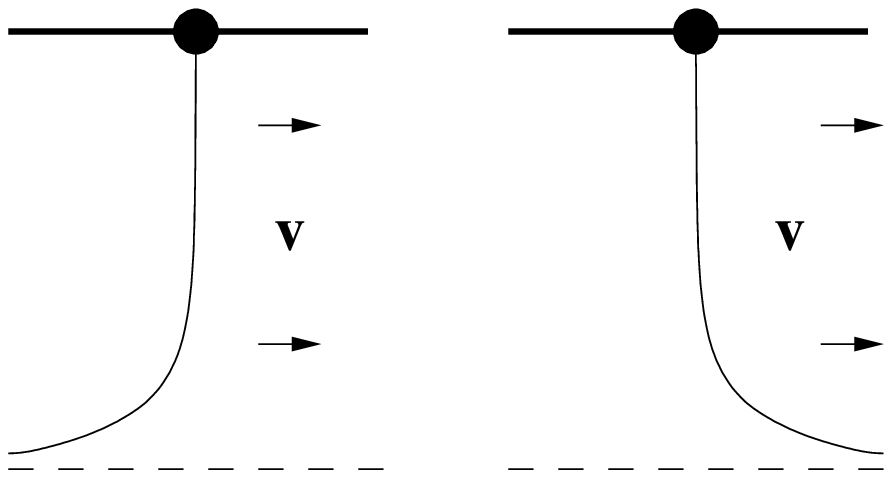,width=2.2in}}
    \caption{Schematic drawing of the physical solution (left)
    in which energy flows toward the horizon,
    and the unphysical energy solution (right)
    in which energy flows away from the horizon.}
\label{twosol}
 }
 \end{FIGURE}

The rate at which energy flows down the string
is given by $\pi^1_t$.
As seen in Eq.~(\ref{eq:pi}), this energy flux is proportional to~$C$.
If $C$ is positive, then
energy flows down the string toward the horizon,
and the string profile resembles a tail being dragged
behind the moving quark,
as illustrated on the left in Figure~\ref{twosol}.
If $C$ is negative, then one has the time-reversed situation:
energy flows upward from the horizon and
the tail of string leads the quark.
We postulate (as in Ref.~\cite{sonstarinets})
that the physical process we want to describe requires us to pick
{\it purely outgoing} boundary conditions at the horizon.%
\footnote
    {
    Here and elsewhere, outgoing refers to moving out of
    the physical region and into the black hole.
    }
We have to discard the time-reversed solution in which
energy flows from the black hole to the moving quark.

The resulting rates at which energy and momentum
flow toward the horizon are
\begin{subequations}
\begin{eqnarray}
\label{Erate}
    \left . \pi^1_t \right |_{u=u_h}
    &=& T_0 L^2 \, u_h^2 \, \frac{v^2}{(1-v^2)^{2/d}}
    \quad \mathop{\longrightarrow}\limits^{d=4} \quad
    \coeff\pi 2 \sqrt{\lambda} \, T^2 \, \frac{v^2}{\sqrt{1-v^2}} \,, \\
\noalign{\hbox{and}}
\label{Prate}
    \left . -\pi^1_x \right|_{u=u_h}
   &=& T_0 L^2 \, u_h^2 \, \frac{v}{(1-v^2)^{2/d}}
    \quad \mathop{\longrightarrow}\limits^{d=4} \quad
    \coeff\pi 2 \sqrt{\lambda} \, T^2 \, \frac{v}{\sqrt{1-v^2}} \,,
\end{eqnarray}
\label{EPrates}
\end{subequations}
respectively.

The stationary solution $x_+(u,t)$ given by
Eqs.~(\ref {eq:analytic}) and (\ref{analyticsol})
describes an open string which runs from the
AdS boundary at $u=\infty$ and asymptotically approaches
the horizon at $u=u_h$.
By truncating the solution at an arbitrary radius $u_m > u_h$,
one may equally well regard it as describing an open string running
from a D7-brane with minimal radius $u_m$ down to the horizon.
The rates (\ref {EPrates}) at which energy and momentum flow down the string
are completely independent of $u_m$.

Standard Neumann boundary conditions would demand that the
momentum flux $\pi_x^1$ vanish at the flavor brane ---
so our solution with a constant non-zero $\pi_x^1$ does not
satisfy Neumann boundary conditions.
The solution is physical, but
there must be a force acting on the string endpoint
and feeding energy and momentum into the string.
A constant electric field on the flavor brane provides precisely such a force.
The field alters the Neumann boundary condition to the force balance
condition $\pi_x^1 =  -F_{tx}$.%
\footnote
    {
    D-branes in string theory naturally support gauge fields living
    on their worldvolume under which the endpoints of strings are
    charged. Turning on the worldvolume gauge field will once more
    change the embedding of the D7-brane and hence the relation
    between $u_m$ and $m$. Since the results we find for the motion
    of the string in the presence of the external field is
    independent of $u_m$ (and hence $m$), the precise relation between
    the two in the presence of the field is not important for our
    purposes. It is, however, worthwhile noting that the electric
    field on a D-brane cannot become larger than a critical value, at
    which point the force pulling the endpoints of a string apart
    due to the field wins against the string tension and the system becomes
    unstable. This happens when $(F_{tx}/T_0)^2 = |g_{tt}
    g_{xx}| = u^4 - u_h^4$. For small mass the brane will extend down
    to a small value of $u$, so there will be a maximum field that
    the brane can support (and hence a maximal velocity) that
    decreases as the quark mass decreases.
    }

Although the flux of energy and momentum in this stationary solution
is finite, the total energy and momentum of the string is infinite,
due to the contribution to the integrals (\ref {EPeqs})
close to the horizon.
If one simply inserts a lower limit $\umin > u_h$,
together with an upper limit equal to the radius $u_m$ of the D7-brane,
then the resulting energy and momentum (in $d=4$) are
\begin{subequations}
\begin{eqnarray}
    E &=&
   - \int_{\umin}^{u_m} du \> \pi^0_t(u)
    =
    T_0 L^2 \, \frac{1}{\sqrt{1-v^2}} \,
    \left[ u_m - \umin + v^2 \Lambda(\umin) \right] ,
\\[5pt]
    p &=&
    \int_{\umin}^{u_m} du \> \pi^0_x(u)
    =
    T_0 L^2 \, \frac{v}{\sqrt{1-v^2}} \,
    \left[ u_m - \umin + \Lambda(\umin) \right] \ ,
\end{eqnarray}
\label{eq:EPstring}
\end{subequations}
where
\begin{equation}
    \Lambda(\umin) \equiv
    \frac{u_h}{4} \!
    \left[\,
     2 \, \arctan \, \frac{u_{min}}{u_h}
    -2 \, \arctan \, \frac{u_{m}}{u_h}
    - \ln \frac{(u_{m}+u_h)(u_{min}-u_h)}{(u_{m}-u_h)(u_{min}+u_h)}
    \right] .
\end{equation}
As $\umin \to u_h$,
this function diverges logarithmically as
$
    -\coeff 14 \, u_h \ln (\umin {-} u_h)
$.
The interpretation of this IR divergence will be discussed momentarily.

\subsubsection*{Field theory interpretation}
\label{sec:ftinterp}

We want to interpret this stationary string solution
as describing the steady-state behavior of a massive quark
moving through the $\Nfour$ plasma under the influence of a constant
electric field%
\footnote
    {
    $\cal E$ is a ``real'' electric field --- a $U(1)$ gauge field
    coupled to quark flavor, having nothing to do with the
    $SU(\Nc)$ gauge fields.
    Such an electric field acts on fundamental representation quarks,
    but has no effect on any of the $\Nfour$ SYM fields.
    }
${\cal E} = F_{tx}$.
The quark velocity will asymptotically approach an equilibrium value $v$
at which the rate of momentum loss to the plasma is balanced by the
driving force exerted by the electric field.
The rate at which the electric field does work on the quark,
namely $v \cdot {\cal E}$, should also equal the rate at which the
quark loses energy to the medium.
The results (\ref {EPrates}), plus the
force balance condition at the string endpoint,
\begin{equation}
    \pi^1_x = -\cal E \,,
\label{eq:balance}
\end{equation}
are completely consistent with this interpretation,
provided one regards energy and momentum flow toward the
horizon as energy and momentum transfer to the thermal medium.

If the quark behaves as an excitation with some effective mass $M$
and momentum $p = M v / \sqrt{1-v^2}$, then
the result (\ref{Prate}) for the momentum transfer rate
is equivalent to a momentum loss rate ${dp}/{dt} = -\mu p$
for the quark, with
\begin{equation}
    \mu M = \coeff \pi 2 \sqrt\lambda T^2 = \pi T \, \Delta m(T) \,.
\label {eq:muM}
\end{equation}
Just as in the toy model discussed in the Introduction,
the momentum flow of our steady state solution
determines $\mu M$, but not $\mu$ or $M$ individually.
Note that $\mu M$ is independent of the flavor brane location
$u_m$, and hence is independent of the physical quark mass.

The energy of the string should be regarded as the total
excess free energy of the system --- 
the free
energy minus its equilibrium value at the given temperature.
In other words, the energy of string includes all the effects
of the disturbance to the $\Nfour$ plasma produced by the
moving quark.
The stationary moving string solution is describing a system
in which a quark has been forcibly dragged through the plasma
(which is infinite in extent) for an unbounded period of time.
The constant rate of work done by the external electric field
thus translates into an infinite input of energy to the plasma.
This unbounded input of energy is the physical origin of the IR
divergence in the string energy,
as may be seen explicitly by noting that the logarithmically
divergent function $\Lambda(u_{\rm min})$  appearing in
the energy and momentum (\ref {eq:EPstring}) is proportional to the
difference in position (in $x$) between the two ends of the cut-off
string,
\begin{eqnarray}
    \Lambda(u_{\rm min})
    &=&
    u_h^2 \, \left| \frac {\Delta x(u_{\rm min})} v \right|
\\
\noalign{\hbox{with}}
    \Delta x(u_{\rm min})
    &\equiv&
    x_+(u_m,t) - x_+(u_{\rm min},t) \,.
\end{eqnarray}
Comparison with Eq.~(\ref {EPrates}) shows that
the cut-off string energy and momentum are just
a boosted static energy plus the net input of energy and momentum
required to move the quark a distance $\Delta x$ at velocity $v$,
\begin{subequations}
\begin{eqnarray}
    E &=& T_0 L^2 \> \frac {(u_m{-}u_{\rm min})}{\sqrt{1-v^2}} \,
    + \frac 1v \, \frac {dE}{dt} \> \Delta x(u_{\rm min}) \,,
\\
    p &=& T_0 L^2 \> \frac {v \, (u_m{-}u_{\rm min})}{\sqrt{1-v^2}}
    + \frac 1v \, \frac {dp}{dt} \> \Delta x(u_{\rm min}) \,,
\end{eqnarray}
\label{eq:EP2}
\end{subequations}
where $dE/dt = \pi^1_t$ and $dp/dt = -\pi^1_x$
are the rates at which the external electric field transfers
energy and momentum to the quark.
Note that the $u_{\rm min} \to u_h$ limit of
$T_0 L^2 (u_m{-}u_{\rm min})$ is just the static rest energy $\Mrest(T)$.

The result (\ref{eq:EP2}) suggests that one might interpret
$\Mrest(T)/\sqrt{1-v^2}$ as the energy
(and $v$ times this as the momentum)
of a quark moving at velocity $v$ through the plasma.
Or equivalently, that the appropriate thermal dispersion relation
is just a relativistic dispersion relation but with mass $\Mrest(T)$.
This, we believe, is too simplistic.
A quark moving through the thermal plasma is a quasiparticle ---
an elementary excitation of the system with a finite thermal width
given by the damping rate $\mu$.
The (free) energy of a quark moving through the medium is only defined
to within an uncertainty given by its thermal width.
A natural operational definition is to start with a static quark,
at rest in the thermal medium, turn on an electric field which
accelerates the quark to the desired velocity in some time $\tau$,
and then define the energy of the quark as its initial rest energy plus
the work done by the electric field while accelerating the quark.
The acceleration time $\tau$ should be small compared to the damping
time $\mu^{-1}$, to avoid counting energy which has
already been lost to the medium.
But $\tau$ should also be large compared to the inverse kinetic
energy of the quark, to minimize the quantum uncertainty in the energy.
Satisfying both conditions is only possible if the thermal width
is small compared to the energy, which is the basic condition
defining a good quasiparticle.
Choosing $\tau \sim (\mu M v^2)^{-1/2}$ (for non-relativistic motion)
balances the two uncertainties and gives a
limiting precision with which one can define
the kinetic energy of a moving quark that scales as $\lambda^{1/4} \, T/p$.
Finding the requisite string solution with such a time-dependent
electric field has not (yet) been done.

\subsection{Quasinormal modes}
\label{sec:qnm}

Instead of considering a quark moving under the influence of an
external electric field, we now turn to the motion of a quark
decelerating in the thermal medium, in the absence of any
external forcing.
The setup here is the analog of the second gedanken experiment for the
toy model discussed in the Introduction.
We will focus on the late-time, and hence low-velocity, behavior.
We extract this late-time dynamics by analyzing
small perturbations to the static string which describes
a quark at rest.
A key ingredient will be the purely outgoing boundary conditions
at the horizon which capture the dissipative nature of the process and
have been shown to reproduce appropriate thermal physics \cite {sonstarinets}.
With these boundary conditions, the problem becomes a quasinormal
mode calculation on the string worldsheet.
To complement the linear analysis of this section,
in Section~\ref{sec:numerics} we will also perform a numerical
analysis of the full, time-dependent problem.

The linearized equation of motion for small fluctuations around the
static straight string, $x(u,t)=x_0$
means treating $\dot x$ and $x'$ as small and retaining only terms
linear in derivatives of $x$.
From Eq.~(\ref{eq:-g}) one sees that this corresponds to
replacing $-g/L^4$ with 1, which reduces
the full string equation of motion (\ref{eomt}) to%
\footnote
    {
    In $d=3$ dimensions, this differential equation
    coincides with the massless scalar wave equation in
    the $AdS_4$ black hole background, at zero spatial momentum.
    In other dimensions, including the $d=4$ case of interest,
    the linearized string equation (\ref{linearizedeomt})
    differs from the scalar wave equation.
    }
\beq
\frac{\partial}{\partial u} \left ( h u^2 x' \right )
= \frac{u^2}{h}\, \ddot{x} \,.
\label{linearizedeomt}
\eeq

To select the physically relevant solution,
we impose purely
outgoing boundary conditions at the horizon.
These boundary conditions make the resulting boundary value problem
non-hermitian and
the resulting quasinormal modes will have real exponential time dependence.
Close to the horizon,
the most general solution of the wave equation (\ref{linearizedeomt})
has the form
\beq
\label{nha}
x(u,t) =
F\Big(t+\frac{1}{u_h d}\log(\epsilon) \Big) +
G\Big(t-\frac{1}{u_h d}\log(\epsilon) \Big) \ .
\eeq
where $\epsilon \equiv u/u_h -1$ [or $u=(1 + \epsilon)u_h $],
and $F(x)$ and $G(x)$ are arbitrary (differentiable) functions.
Purely outgoing means that $G=0$ in this regime.

Specializing to $e^{-\decay t}$ time dependence and
introducing, for convenience, a dimensionless radial coordinate
$y \equiv u/u_h$ and dimensionless decay rate $\wn = \decay/u_h$,
the linearized wave equation (\ref{linearizedeomt}) becomes the
eigenvalue equation
\begin{subequations}
\label{yequation}
\begin{eqnarray}
    &&\mathcal L \, x = \wn^2 \, x \,,
\\
\noalign{\hbox{with}}
    &&\mathcal L \equiv
    (1-y^{-d}) \, \frac {d}{dy} \, y^4 (1-y^{-d})\, \frac{d}{dy} \,.
\end{eqnarray}
\end{subequations}
Imposing outgoing boundary conditions at the horizon
and Neumann boundary conditions at the flavor brane
leads to a discrete spectrum of quasinormal mode decay rates.

Close to the horizon at $y=1$,
the purely outgoing solution to Eq.~(\ref{yequation})
is proportional to $(y{-}1)^{ - \wn/d}$ with $\wn$ positive.
This diverges as $y \to 1$,
showing that there are non-uniformities between the
large time and near horizon limits.
In particular, the assumption that $\sqrt{-g}/L^2 \approx 1$
is only valid at sufficiently late times,
when $t + (u_h d)^{-1} \log(y{-}1)$ is sufficiently large.
To evaluate the deviation of $\sqrt{-g}/L^2$ from unity,
one needs the next-to-leading term in the near horizon asymptotics.
We find
$
x(y,t) = A (y{-}1)^{-\wn/d} e^{-\wn u_h t}
\left\{ 1 + \wn B(y{-}1)  + \mathcal O[(y{-}1)^2] \right\}
$
with
$
B = \frac{3-d}{2d} + \frac{2}{d-2\wn}
$.
This gives
$$
\frac{-g}{L^4}  = 1 -
\frac{4 A^2 \mu^2}{d{-}2\wn} \, (y{-}1)^{-2\wn/d} \, e^{-2 \mu t}
\times [ 1 + {\mathcal O}(y{-}1) ] \,,
$$
which shows that, at any fixed position outside the horizon,
$\sqrt{-g}/L^2 \approx 1$ for sufficiently large times.

The near horizon asymptotics are useful also for exploring the
small mass limit $y_m \to 1$ of our quasinormal mode problem.
It is possible for $x(y,t)$ to obey Neumann boundary conditions at $y=y_m$,
in the limit $y_m \to 1$,
if $\gamma = \coeff d2 + \mathcal O(y_m{-}1)$.
This enables the
first two terms in the asymptotic series expansion
for $x(y,t)$ to give comparable, and canceling, contributions to the slope
$x'(y_m,t)$.
This result, $\gamma \to \coeff d2$,
gives the intercepts of the $d=4$ and $d=2$ curves
shown below in Figure \ref{qnm}.
Converting from the dimensionless decay rate $\gamma$
back to $\mu$ yields the result, valid for all $d$,
that $\mu = 2 \pi T$ in the limit where the flavor brane approaches the horizon.

\subsubsection*{Large mass limit}
\label{sec:smallw}

The differential equation (\ref{yequation}) does not appear to have
a simple solution for arbitrary $d$.
In the special case of $d=2$, the
differential equation can be solved in terms of associated Legendre functions,
as discussed in Appendix~\ref{app:qnm}.
For the $d=4$ case of interest, Eq.~(\ref{yequation}) is
a particular example of the Heun equation, a differential equation
with four regular singular points.
Heun functions are difficult to work with, and other values of $d$
appear to be even more difficult to treat analytically.

In the absence of a simple analytic solution to Eq.~(\ref{yequation}),
we attempt a power series solution in $\wn$,
\be
x(y) = x_0(y) + \wn \, x_1(y) + \wn^2 \, x_2(y) + \cdots \,.
\ee
We will focus on the large mass regime where the flavor
brane position $y_m \equiv u_m/u_h \gg 1$, and
we will find an iterative solution where $\wn = \mathcal O(1/y_m)$.
For the moment, the correlation between $\wn$ and $y_m$
 may be viewed as an assumption which will be
verified {\it a posteriori}.
Requiring $\mathcal L \, x = \wn^2 \, x$ implies that
\be
    {\mathcal L}\, x_0 = 0 \,,\quad
    {\mathcal L}\, x_1 = 0 \,,\quad \mbox{and} \quad
    {\mathcal L}\, x_2 = x_0 \,.
\ee
We want to satisfy Neumann boundary conditions at the flavor brane,
$x'(y_m) = 0$,
together with outgoing boundary conditions at the horizon.
As seen above, this requires that $x(y) \sim A \, (y{-}1)^{-\wn/d}$ as $y\to 1$,
for some constant $A$,
or equivalently
\be
    x'(y) \sim -A \, \frac{\wn}{d} \, (y{-}1)^{-1-\wn/d}
    =
    A \left[
    -\frac{\wn}{d} \, \frac 1{y{-}1}
    + \frac{\wn^2}{d^2} \, \frac{\ln(y{-}1)}{y{-}1}
    + \mathcal O(\wn^3)
    \right] .
\label{eq:outgoing-exp}
\ee
The constant $x_0(y) = A$ is
the only homogeneous solution which obeys Neumann boundary
conditions at the flavor brane at $y=y_m$.
To generate the $\mathcal O(\wn)$ term in the near horizon behavior
(\ref{eq:outgoing-exp}), the first order correction
$x_1(y)$ must equal the second homogeneous solution of $\mathcal L \, x = 0$.
The derivative of this homogeneous solution
(which will be sufficient for our purposes) is
\begin{equation}
    x_1'(y) = -A \, \frac {y^{d-4}}{y^d{-}1} \,.
\label{eq:x1'big}
\end{equation}
At the flavor brane, $x_1'(y_m) = -A \, y_m^{-4} + \mathcal O(y_m^{-4-d})$.
This violates the Neumann boundary conditions.
However, if $\wn = \mathcal O(1/y_m)$, then this
violation can, and will, be compensated by the next order term.
Moving on to $x_2(y)$ and solving $\mathcal L \, x_2 = x_0$ gives
\be
x_2'(y) =
    A \, \frac{y^{d-3}}{y^{d}{-}1}
    \left[
    1- {}_2 F_1\left(\coeff{1}{d}, 1, \coeff{d+1}{d}; y^{d} \right)
    \right]
    +
    C \, \frac {y^{d-4}}{y^d{-}1} \,.
\label{eq:x2'}
\ee
As $y \to 1$, the leading term in the hypergeometric function is
\be
{}_2 F_1 \left(\coeff{1}{d}, 1, \coeff{d+1}{d}; y^{d} \right)
= - \frac{1}{d}\, \ln (y{-}1) + {\mathcal O}(1) \ ,
\ee
and so
\be
    x_2'(y) =
    \frac{A}{d^2} \frac{1}{y{-}1} \Big[ \ln (y{-}1) + \mathcal O(1) \Big] .
\label{nhtwo}
\ee
The logarithmic term is precisely what is required to generate
the $\mathcal O(\wn^2)$ term in
the outgoing boundary condition (\ref {eq:outgoing-exp}).
The unwanted non-logarithmic $(y{-}1)^{-1}$ term is eliminated
by an appropriate choice for the coefficient $C$ of the homogeneous
term in (\ref{eq:x2'}).

When evaluated at $y_m \gg 1$,
the hypergeometric function is negligible compared to 1,
\be
{}_2 F_1\left(\coeff{1}{d}, 1, \coeff{d+1}{d}; y_m^{d} \right) = {\mathcal O}(1/y_m) \,,
\ee
so that
\begin{equation}
    x_2'(y_m) = A \, y_m^{-3} + \mathcal O(y_m^{-4}) \,.
\label{eq:x2'big}
\end{equation}
The Neumann boundary condition requires
\begin{equation}
    0 = x'(y_m) =
    \wn \, x_1'(y_m) +
    \wn^2 \, x_2'(y_m) +
    \cdots \,.
\end{equation}
Inserting the explicit forms (\ref{eq:x1'big}) and (\ref{eq:x2'big}),
one sees that the leading $\wn /y_m^4$ term from $x_1'(y_m)$
will cancel the leading $\wn^2 / y_m^3$ term from $x_2'(y_m)$ provided
\be
    \wn = \frac{1}{y_m} + {\mathcal O}\left( 1/{y_m^{2}} \right)  .
\label{qnmresult}
\ee
This value of $\wn^2$ is the smallest eigenvalue of
$\cal L$ (for the given boundary conditions).
All other eigenvalues are $\mathcal O(1)$ as $y_m \to \infty$.
Expressing the result (\ref{qnmresult}) in terms of the original variables,
the lowest quasinormal mode decay rate is
\beq
 \decay = \frac{u_h^2}{u_m} + \mathcal O(u_h^3/u_m^2) \,.
\label{eq:Gamma}
\eeq

The motion of the string endpoint in this
quasinormal mode is a simple exponential,
\beq
x(u_m,t) -x_0 \propto A\, e^{- \decay t} \,,
 \eeq
so the velocity of the string endpoint satisfies
\be
\dot v = - \decay \, v \,.
\ee
Inserting the result (\ref{eq:Gamma}) for $\decay$
and using the large mass limit of the relation (\ref {eq:um/m})
between the brane position and quark mass,
namely $u_m = m/(T_0 L^2)$, converts this result to
\beq
\frac{d p}{dt} = - (T_0 L^2 \, u_h^2) \, v \,,
\eeq
with $p = mv$.  This agrees perfectly
with our earlier result (\ref{Prate}) at small $v$.

\subsubsection*{Arbitrary mass}

\begin{FIGURE}[t]
    {
    \centerline{\psfig{figure=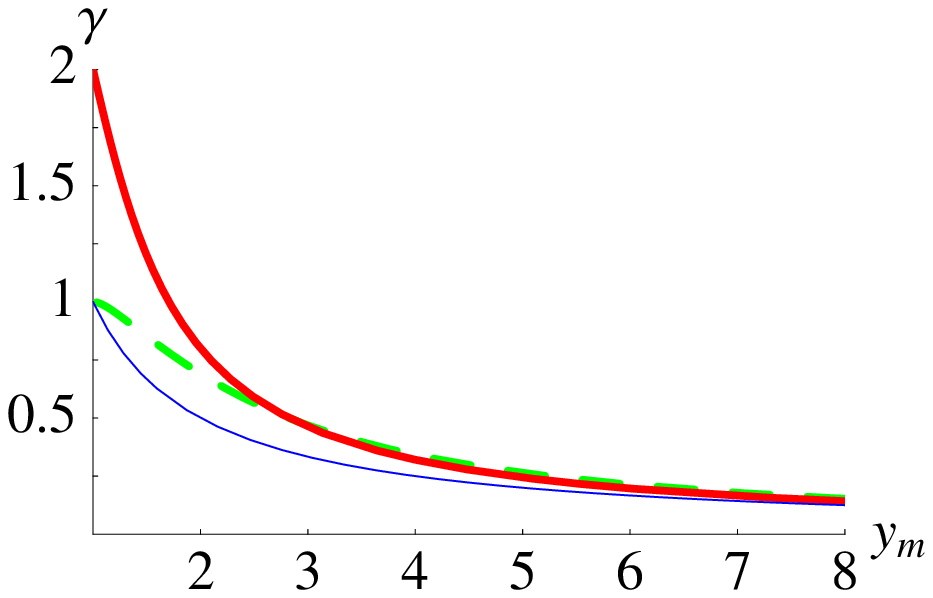,width=2.8in}}
    \caption
    {
    Lowest quasinormal mode decay rate $\wn$ as
    a function of mass parameter $y_m$,
    for $d=2$ (green dashed line) and
    $d=4$ (thick red solid line),
    together with the leading large mass
    analytic form $\wn=\frac{1}{y_m}$ (thin blue line)
     }
    \label{qnm}
    }
 \end{FIGURE}

To find the lowest eigenvalue of the quasinormal mode operator $\mathcal L$
for an arbitrary D7-brane location $y_m$, it is necessary to resort to
numerical analysis.
A simple shooting algorithm suffices.
At a point close to the horizon,
$y = 1+\epsilon$ with $\epsilon \ll 1$,
one sets $x(1{+}\epsilon)=1$ and
$x'(1{+}\epsilon) = - \frac{\wn}{d} \frac{1}{\epsilon}$.
This enforces the outgoing boundary condition.
Then,
for various values of $\wn$,
one integrates the differential equation
$(\mathcal  L - \wn^2) \, x = 0$ out to the flavor brane,
and successively refines $\wn$ to locate the minimal
value that satisfies the Neumann boundary condition $x'(y_m) = 0$.

In Figure~\ref{qnm}
we plot the resulting lowest quasinormal decay rate, as a function
of the flavor brane location $y_m$,
in dimensions $d=2$ and 4.
These numerical values are consistent with our large mass result
that $\wn=y_m^{-1} + \mathcal O(y_m^{-2})$.
In Appendix~\ref{app:qnm},
we show that the $d=2$ case is analytically soluble,
and the resulting frequencies satisfy the equation
\be
y_m = \frac{1}{\wn} - \frac{\wn}{2} + {\mathcal O}(\wn^2) \ .
\ee

When we compare the actual shape $x(u,t)$ of the QNM with the
analytic stationary
solution, we find that they agree when $\gamma \ll 1$ (that is for large mass),
which is when the external field needed to maintain the velocity is small.

\subsubsection*{Low velocity dispersion relation}

In addition to extracting the lowest quasinormal mode decay rate,
the linearized equation of motion (\ref {linearizedeomt})
may also be used to find the dispersion relation of a
quark moving at low velocity.
If $\dot x$ and $x'$ are small, so that $\sqrt{-g/L^4} \approx 1$,
then the momentum density
$
    \pi_x^0 = T_0 L^2 \, u^2 \, h^{-1} \, \dot x
$.
For a quasinormal mode with
exponential time dependence,
$x(u,t) = x(u) \, e^{-\decay t}$,
the momentum density may be rewritten as
\be
\pi_x^0
= -\frac {T_0 L^2}{\decay} \> u^2 \, h^{-1} \> \ddot x
= -\frac {T_0 L^2}{\decay} \> (h u^2 x')' \,,
\ee
where the last form follows from
the linearized equation of motion (\ref{linearizedeomt}).
This current is easily integrated to find the total momentum
carried by the portion of the string running from the flavor brane
down to an IR cutoff $u_{\rm min} > u_h$,
\be
    p
    = \int_{u_{\rm min}}^{u_m} du \> \pi_x^0
    = \frac{T_0 L^2}{\decay} \> h u^2 x' \Big|^{u=u_{m}}_{u=u_{\rm min}} \ .
\ee
Because of the Neumann boundary conditions, the upper endpoint does
not contribute.
Hence
\be
    p = \frac{T_0 L^2 }{\decay} \>
    h(u_{\rm min}) \, u_{\rm min}^2 \, x'(u_{\rm min}) \,.
\label{eq:qnmmomentum}
\ee

The energy may be evaluated similarly.
We are interested in the deviation of the energy from
the static rest energy, and it is therefore necessary to
keep all terms up to quadratic order in $\dot x$ and $x'$ in
the energy density $\pi_t^0$.
Suitably expanding $\sqrt{-g/L^4}$,
the energy density is
\be
    \pi_t^0 =
    -T_0 L^2 \Big[ 1
    + \half h u^2 (x')^2
    + \half \decay^2 \, h^{-1} u^2 x^2
    + \mathcal O(x^4)
    \Big] \,.
\ee
Using the linearized equation of motion (\ref{linearizedeomt}),
one may express the resulting energy solely as endpoint contributions,
\be
    E = -\int_{u_{\rm min}}^{u_m} du \> \pi_t^0 = T_0 L^2
    \Big[
    (u_m - u_{\rm min}) -
    \half h(u_{\rm min}) \, u_{\rm min}^2 \,
    x(u_{\rm min}) \, x'(u_{\rm min})
    \Big] ,
\label{eq:qnmenergy}
\ee
where Neumann boundary conditions have again caused the non-static
boundary term at $u_m$ to vanish.

Taking $u_{min}$ to be close to the horizon $u_h$,
and inserting the outgoing near-horizon behavior
\be
x(u) \sim A (u-u_h)^{-\decay/d u_h} \, e^{-\decay t} \ .
\ee
into expressions
(\ref{eq:qnmmomentum}) and (\ref{eq:qnmenergy}), yields
a simple relation between the energy and momentum,
\be
E = T_0 L^2 (u_m {-} u_{\rm min})
+ \frac{p^2}{2 \Mkin} \,,
\label {eq:qnmE}
\ee
where the ``kinetic mass''
\be
    \Mkin \equiv \frac {T_0 L^2 \, u_h^2}{\decay}
    = \frac{ \pi \sqrt{\lambda} T^2}{2\decay}
    = \frac {\pi T}{\decay} \, \Delta m(T)
    \,.
\label {eq:Mkin}
\ee
This kinetic mass is independent of the IR cutoff $u_{\rm min}$,
while the $u_{\rm min} \to u_h$ limit of
the first term of the energy (\ref{eq:qnmE})
is just the thermal rest energy $\Mrest = T_0 L^2 (u_m {-} u_h)$.

In the heavy mass limit,
the decay rate $\decay$ approaches
$u_h^2 / u_m = \pi T \, \Delta m(T)/m$
[see Eqs.~(\ref{eq:Gamma}) and (\ref{eq:um/m})].
Therefore, in this limit the ratio of the kinetic mass $\Mkin$
(or the thermal rest mass $\Mrest$) to the Lagrangian mass $m$
approaches one.
For smaller $u_m$, corresponding to order one values of $m/\Delta m(T)$,
there is a more complicated relationship between the kinetic mass
and the Lagrangian mass.
The kinetic mass $\Mkin$ is plotted as a function of
$m$ in Figure~\ref{fig:masses}.

Finally,
note that the $e^{-\mu t}$ time dependence of the lowest quasinormal mode
plus the value (\ref{eq:Mkin}) for the kinetic mass
are equivalent to viscous drag,
\be
\label{resqnm}
\frac{dp}{dt} = -\mu \, p \,,
\ee
with a friction coefficient
\be
    \mu
    = \frac \pi 2 \, \frac{\sqrt{\lambda} T^2}{\Mkin}
    = \pi T \, \frac{\Delta m(T)}{\Mkin} \,,
\ee
which for $M=\Mkin$ is completely consistent with the value (\ref {eq:muM})
of $\mu M$ extracted from the analytic stationary solution.

\section{Quark-antiquark solutions}
\label{sec:numerics}

Thus far, we have deduced $\mu M$ (or the viscous drag as a function of
velocity) from the stationary analytic solution, whose analysis was
valid for all velocities.
And we have deduced the value of the friction coefficient $\mu$ itself
from the linearized, quasinormal mode analysis, valid for late times
and hence small velocities.
Both approaches reveal IR sensitivity of the total string
energy and momentum, which we argue should be viewed as reflecting
an unavoidable level of arbitrariness in defining the partitioning
of the total system energy (or momentum) into a piece associated
with the moving quark plus a remainder associated with the long
disturbance in the medium.

A more physical approach for dealing with this IR sensitivity
is to change the question.
Instead of considering a single quark moving through the plasma,
one may study $q\bar q$ pair creation ---
that is, the dynamics of a quark-antiquark pair where the
quark and antiquark are initially flying apart from each other.
Such a pair corresponds to a string with both endpoints on the D7-brane.
The previous IR sensitivity due to string dynamics arbitrarily close to
the horizon will be cut-off by the
finite quark-antiquark separation, which will limit how far down toward
the horizon the middle of the string can ``sag''.

For simplicity, we will limit our attention to the case of
back-to-back motion, so the total momentum will vanish
and the entire string worldsheet will lie within the three-dimensional
$(t,u,x)$ slice of the AdS-black hole geometry.
Hence,
we need to find non-stationary solutions of
the partial differential equation (\ref {eomt})
with physically relevant initial conditions.
To do so, we need time dependent numerics.

To set up the numerical problem,
it is convenient to remap the infinite range of the radial coordinate
$u \in (u_h,\infty)$ onto a finite interval.
To do so, we define
\be
    z = \frac 1y = \frac {u_h}{u} \,.
\ee
We will specialize to $d=4$ and choose units where $u_h = 1$,
so the line element of the black D3-brane gravitational background becomes
\be
ds^2 =
\frac{L^2}{z^2} \left( -f(z) \, dt^2 + d\vec x^2 + \frac{dz^2}{f(z)} \right) ,
\ee
where $f(z) \equiv 1-z^4$ and $\vec x = (x^1, x^2, x^3)$.
Temperature dependence may be restored later by rescaling
the coordinates: $t \to u_h \, t$, $x^i \to u_h \, x^i$. 
In this coordinate system, the black hole horizon is located at $z=1$ and the
$AdS$ boundary at $z=0$.  Our open string will end on a D7-brane which fills
the five dimensional space from $z=0$ to $z=z_m$.
We will assume that both string endpoints are located at $z=z_m$,
and that the string extends only in the $z$ and $x=x^1$ directions.

    Changing variables from $u$ to $z$ in
the 1+1 dimensional partial differential equation (\ref{eomt}),
and then discretizing on a rectangular grid%
\footnote
    {
    As is done internally in canned PDE solvers
    such as Mathematica's {\tt NDSolve}.
    }
in $z$ and $t$
turns out to be a bad approach.
Numerical stability rapidly degrades as the
string endpoints separate and the middle of the string
gets closer to the horizon.
The net result is that the numerical integration breaks down
after a very limited amount of time.

We have found that a much better starting point for numerical integration is
the Polyakov action with a worldsheet metric that is a
generalization of conformal gauge.
Recall that the Polyakov action for the string takes
the form
\be
S_{\rm P} = -\frac{T_0L^2}{2}  \int d \sigma \,  d \tau \> \eta^{\alpha \beta}
G_{\mu\nu} \, \partial_\alpha X^\mu \, \partial_\beta X^\nu \, \sqrt{-\eta} \ .
\ee
Here $X(\sigma, \tau)$ is a map from the string world-sheet into space-time,
$\eta_{\alpha \beta}$ is the world-sheet metric,
$G_{\mu\nu}$ is the space-time metric,
$T_0$ is the string tension, and $\sqrt{-\eta}$ is minus the square root of the
determinant of $\eta_{\alpha\beta}$.  We take $0\leq \sigma \leq \pi$.

 From the action $S_{\rm P}$, one derives the usual equations of motion
 for the string,
\begin{align}
\half \eta^{\alpha \beta} \, \frac{\partial G_{\nu\rho}}{\partial X^\mu} \>
\partial_\alpha X^\nu \, \partial_\beta X^\rho \, \sqrt{-\eta}
&=
\partial_\tau \left[ G_{\mu\nu}\sqrt{-\eta} \left( \eta^{\tau\tau} \dot X^\nu + \eta^{\tau\sigma} {X'}^\nu \right) \right]
\nonumber\\
&+ \partial_\sigma \left[G_{\mu\nu}\sqrt{-\eta} \left(  \eta^{\tau \sigma} \dot X^\nu + \eta^{\sigma \sigma} {X'}^\nu \right) \right] ,
\end{align}
along with a constraint on the world-sheet metric,
\be
G_{\mu\nu} \, \partial_\alpha X^\mu \, \partial_\beta X^\nu =
\half \eta_{\alpha \beta} \, \eta^{\gamma \delta} \,
G_{\mu\nu} \, \partial_\gamma X^\mu \, \partial_\delta X^\nu \,,
\ee
produced by the variation of $S_{\rm P}$ with respect to $\eta_{\alpha\beta}$.
The world-sheet metric
$\eta_{\alpha\beta}$ may be integrated out (classically),
converting the Polyakov action into the original Nambu-Goto action.

A standard choice of world-sheet metric is ``conformal gauge'',
in which one chooses the metric to differ from a flat metric
just by an overall conformal factor which is a
function of $\sigma$ and $\tau$,
\be
\| \eta_{\alpha\beta} \| = \left(
\begin{array}{rc}
-1 & 0 \\
0 & 1
\end{array}
\right) e^{\omega(\sigma, \tau)} \ .
\ee
Through trial and error, we have found that this choice also
introduces problems for the numerical integration.
The portion of the world-sheet close to the horizon evolves to late
times far faster
than the portion of the world-sheet closer to the boundary, introducing
large gradients for the embedding
$X(\sigma, \tau)$.
A simple generalization of conformal gauge eliminates this problem
and introduces an extra degree
of freedom which may be tweaked to optimize the performance of the numerical
integrator.
Specifically, we choose a world-sheet metric of the form
\be
\| \eta_{\alpha\beta} \| = \left(
\begin{array}{cc}
-s(\sigma, \tau) & 0 \\
0 & {s(\sigma,\tau)^{-1}}
\end{array}
\right) e^{\omega(\sigma, \tau)} \ .
\ee
We will refer to $s$ as the stretching factor.
With this choice of metric and an arbitrary stretching factor,
the equations of motion become
\begin{subequations}
\begin{eqnarray}
\partial_\tau \!\left( \frac{f \dot t}{s z^2} \right) -\partial_\sigma \!\left( \frac{sf t'}{z^2} \right) & = & 0 \ , \\
\partial_\tau \!\left( \frac{\dot x}{s z^2} \right) - \partial_\sigma \!\left( \frac{sx'}{z^2} \right) & = & 0  \ ,\\
\partial_\tau \!\left(\frac{\dot z}{sfz^2} \right) - \partial_\sigma \!\left( \frac{sz'}{fz^2} \right) &=&
-\frac 1{2s} \left[ \big( {\dot t^2} - s^2 {t'}^2 \big)\, \partial_z \!\left( \frac{f}{z^2} \right)  -
\big({\dot z^2}-s^2 {z'}^2 \big)\, \partial_z \!\left( \frac{1}{fz^2} \right)
\right.
\nonumber
\\
&& \kern 1.7in {} -
\left. \big({\dot x^2}-s^2 {x'}^2 \big)\, \partial_z \!\left( \frac{1}{z^2} \right)
\right] \!.
\end{eqnarray}
\label{numeom}
\end{subequations}
and the constraints, written explicitly, are
\begin{subequations}
\begin{eqnarray}
    0 & = &
    -f \, \dot t t' + \dot x x' + f^{-1} \,  \dot z z' \,,
\\
    0 &=&
    -f \, \big( \dot t^{\,2} +s^2 \, {t'}^2\big)
    + \big( \dot x^2 + s^2 {x'}^2 \big)
    + f^{-1} \big(\dot z^2 + s^2\, {z'}^2 \big) \,.
\end{eqnarray}
\label{constraints}
\end{subequations}
Here (and below), $\dot x \equiv \partial_\tau \, x$
and $x' \equiv \partial_\sigma \, x$, {\em etc}.

The next step in setting up the numerical integration is to find good
initial conditions, a task which is made harder by the constraint
equations.
We have found two consistent sets of useful initial conditions, both
inspired by the classical limit of the leading Regge trajectory of the
string quantized in flat space.  Recall that the leading Regge trajectory
for an open string in flat space with pure Neumann boundary conditions
has a classical limit,
\begin{eqnarray}
t &=& A \, \tau \,, \quad
x = A \cos \sigma \, \sin \tau \,, \quad
z = A \cos \sigma \, \cos \tau \,,
\end{eqnarray}
which corresponds to a line segment of length $A$ spinning in a circle in the $xz$-plane.
Once we introduce a D7-brane along $z=z_0$, there is a closely related classical string state
with mixed Neumann-Dirichlet boundary conditions which serves as our inspiration for initial conditions,
\begin{eqnarray} \label{modRegge}
t &=& A \, \tau \,, \quad
x = A \cos \sigma \, \sin \tau \,, \quad
z = z_0 + A \sin \sigma \, \sin \tau \,.
\end{eqnarray}
This describes a semi-circle expanding and contracting in the $xz$-plane.

Our first set of initial conditions for the AdS-black brane geometry
can be thought of as the $\tau=0$ limit of the semi-circle solution
with some scaling factors that compensate for the fact that we are no
longer in flat space,
\begin{subequations}
\begin{eqnarray}
t(\sigma, 0) = 0 \,, &\quad& \dot t(\sigma,0) = A\, [{1{-}z_m^4}]^{-1/2}  \,,
\\
x(\sigma, 0) = 0 \,, &\quad& \dot x(\sigma, 0) = A\, \cos \sigma \,,
\\
z(\sigma, 0) = z_m \,, &\quad& \dot z(\sigma, 0) = A\, [1{-}z_m^4]^{1/2}\, \sin\sigma \,.
\end{eqnarray}
\label{pointlike}
\end{subequations}
We will call these boundary conditions ``point-like'';
at time $t=0$, the string is
mapped onto a single point in space-time.
The parameter $z_m$, which controls the AdS radius of the string endpoints,
is determined by the quark mass, $z_m \equiv u_h/u_m$.
Notice that the initial speed of the ends of the string
(given by $\dot x/\dot t$) is constrained to be $\sqrt{1-z_m^4}$,
so the initial speed is an increasing function of quark mass.
The ``amplitude'' $A$ controls how much energy is contained in the
initial (zero length) string.
[The explicit relation is given below in Eq.~(\ref{eq:Epoint}).]
Physically, these initial conditions should
resemble the effect of a local current which produces a quark-antiquark
pair when acting on the thermal equilibrium state,
with the quarks having sufficient energy so that their dynamics
may be regarded as classical.

Our second set of initial conditions is an expanding semi-circle
characterized by an adjustable speed $v$.
We take
\begin{subequations}
\label{semicircle}
\begin{eqnarray}
t(\sigma, 0) &=& 0 \,,  \\
x(\sigma,0) &=& A \cos \sigma \,,\\
z(\sigma, 0) &=& z_m + A \, \sin \sigma \,, \\
\dot x(\sigma, 0) &=& v \, \cos \sigma \> \dot t(\sigma, 0) \,.
\\
\noalign{\hbox{The constraint equations then force}}
\dot z(\sigma,0) &=& v \, f \, \sin\sigma \> \dot t(\sigma,0) \,,
\\
\noalign{\hbox{and}}
\dot t(\sigma,0) &=& \frac{A \, s \left[ \sin^2 \sigma + f^{-1} \cos^2 \sigma \right]^{1/2}}{
\left[f-v^2 (\cos^2\sigma + f \sin^2 \sigma) \right]^{1/2}} \,,
\end{eqnarray}
\end{subequations}
with $f$ evaluated at $z(\sigma,0)$.
For these initial conditions to result in a real valued $X(\sigma, \tau)$,
the inequality
\begin{equation}
    v^2 (\cos^2 \sigma + f \sin^2 \sigma) < f
\label{eq:maxv}
\end{equation}
must be satisfied for all $\sigma$.
This macroscopic ``semi-circle'' initial configuration does not correspond
to the action of any local operator,
but the finite size of the string
allows more freedom in choosing the initial speed $v$ of the quarks.

We used the {\tt NDSolve} routine in Mathematica for numerical integration
\cite{Mathematica}. See Appendix \ref{app:error}
for a discussion of numerical error.

\subsection{Forced motion}
\label{sec:analyticconf}

We would like to confirm that the analytic solution presented in
Section~\ref{sec:analytic} is physically relevant.
One might worry
because of the IR divergence in the string energy.
Therefore, we will investigate a quark-antiquark pair
in the presence of a constant electric field,
which will naturally drive the quark and antiquark in opposite directions.

A straightforward numerical integration of the equations of motion
(\ref{numeom}) cannot handle sending the flavor brane all the way
to $z_m = 0$ (or the quark mass to infinity) due to the divergence of $1/f$
at the boundary.
So we will choose positive values of $z_m$ and use
either the point-like initial conditions (\ref{pointlike})
or the semi-circle initial conditions (\ref{semicircle}) to
create a separating quark-antiquark pair at time $t=0$.
Instead of the usual Neumann boundary conditions at $z = z_m$,
we simply fix the endpoint velocity,
$\dot x/\dot t = v$ at $\sigma = 0$ and $-v$ and $\sigma = \pi$.
This corresponds to a time-dependent electric field which
is asymptotically constant,
and whose strength is adjusted precisely to cancel the viscous drag
on the quarks at all times.
As noted above, the speed $v$ must equal $\sqrt{1-z_m^4}$
for point-like initial conditions, or satisfy condition (\ref {eq:maxv}) for
semi-circle initial conditions.
In either case, we find numerical solutions which nicely match
onto two copies of the analytic solution (\ref{analyticsol}) at late times.
As the ends of the string separate,
the middle of the string droops toward the horizon
and the time dependence approaches the stationary form (\ref{analconstantt}).

In Figure~\ref{string}, we plot numeric results for point-like initial
conditions with a D7-brane at $z_m=0.75$ and $A=0.25$.
This value of $z_m$ corresponds to a mass ratio
$m/\Delta m(T) = 1.28$ and a speed $v/c = 0.83$.
As time goes by,
the expanding string, plotted in black, matches onto the analytic
solution, plotted in red, more and more closely.  By $t=1.8$, the two
are practically indistinguishable for $x>0$.

\begin{FIGURE}[t]
{
   \centerline{
   \raisebox{3cm}{(a)}
   \psfig{figure=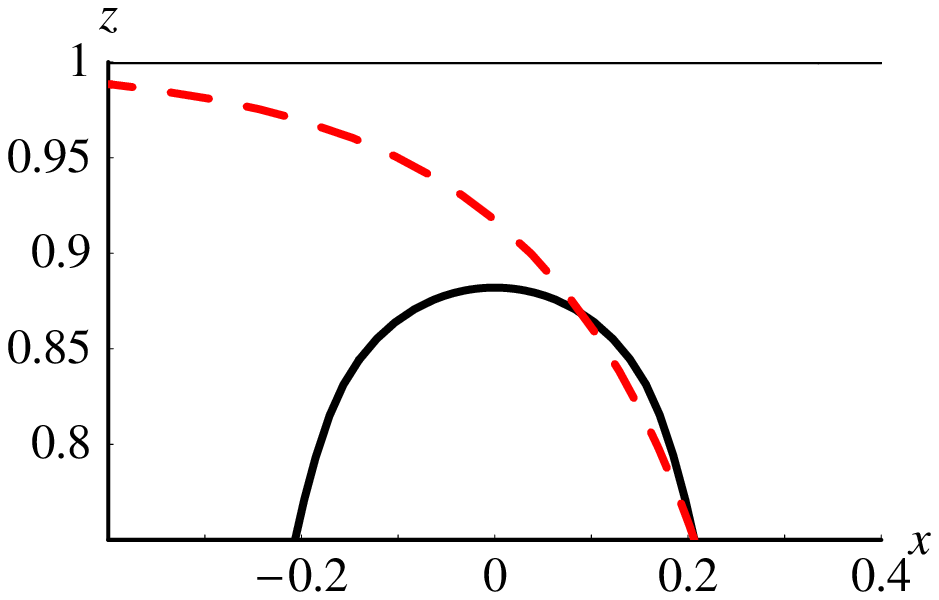,width=2.5in}
   \quad
   \raisebox{3cm}{(b)}
   \psfig{figure=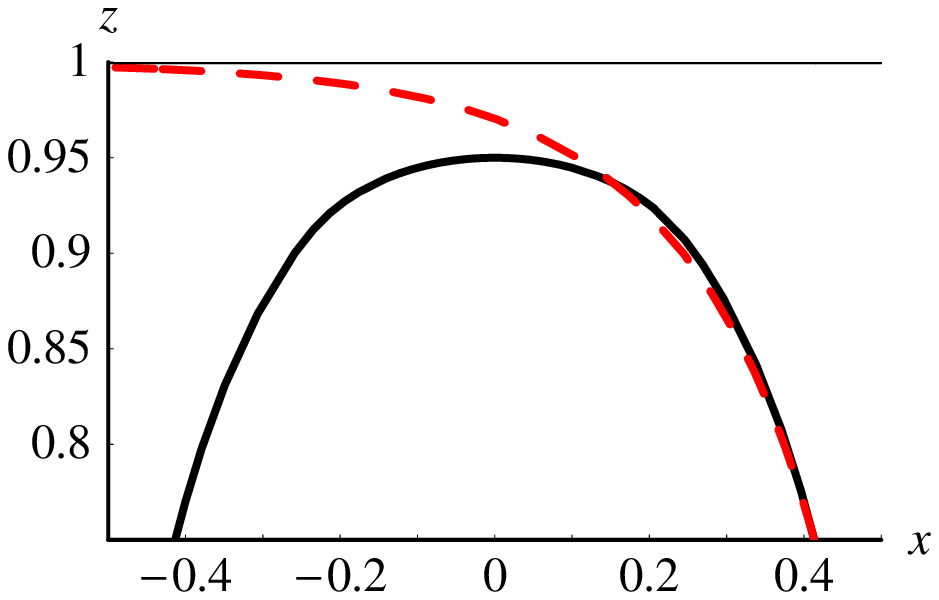,width=2.5in}
   }
   \centerline{
   \raisebox{3cm}{(c)}
   \psfig{figure=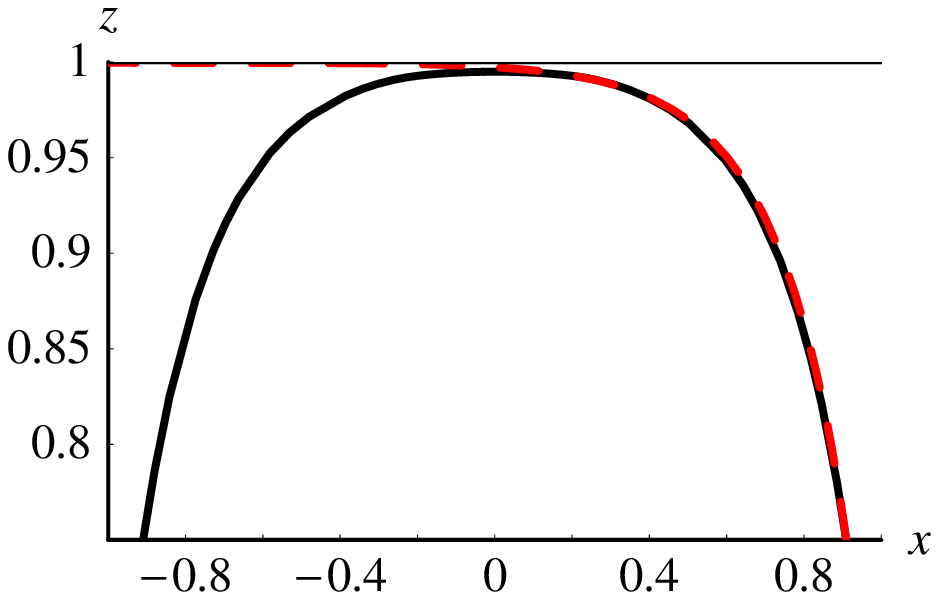,width=2.5in}
   \quad
   \raisebox{3cm}{(d)}
   \psfig{figure=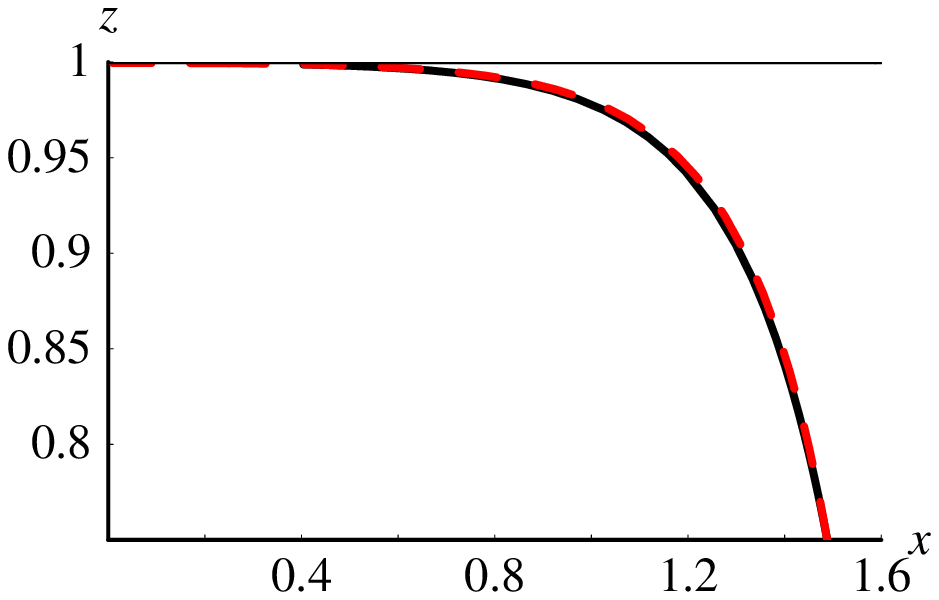,width=2.5in}
   }
    \caption
    {Cross sections of the string world sheet at times (a) $t=0.25$, (b)
    $t=0.5$, (c) $t=1.1$ and (d) $t=1.8$ (in units where $u_h = \pi T = 1$),
    for point-like initial conditions,
    constant velocity boundary conditions,
    and a D7-brane at $z_m = 0.75$.
    The velocity $v/c = 0.83$ and the value of $z_m$
    corresponds to $m/\Delta m(T) = 1.28$.
    The red dashed line shows the constant velocity analytic single
    quark solution.}
\label{string}
}
\end{FIGURE}

Through experimentation, we found that a
stretching function of
\be
s = \frac{1{-}z}{1-3.5 \, z (1{-}z)}
\ee
works particularly well for this string.  The $1{-}z$ factor in the
numerator prevents the string worldsheet from being dragged to late times
close to the horizon.  The denominator ameliorates some distortion of the
constant $\sigma$ and $\tau$ contours of the worldsheet for intermediate
values of $z$.

\subsection{Unforced motion}
\label{sec:qnmconf}

In this subsection 
we consider the motion of a quark-antiquark pair created with point-like
initial conditions (\ref{pointlike}) in the absence of any external forcing.
That is,
we impose the usual Neumann boundary conditions along
the D7-brane.
A short calculation shows that
the energy [given by Eq.~(\ref{EPeqs})]
of our point-like initial configuration is%
\footnote
    {
    The current $\pi^0_t$ can be obtained from a limit of (\ref{raw0current}),
    or more directly from the Polyakov action.
    One finds
    $
    \pi^0_t = -T_0 L^2 \, \eta^{\tau\tau} \sqrt{-\eta} \, G_{tt} \, \dot t \ .
    $
     }
\be
E = T_0 L^2 \, \frac{\sqrt{1-z_m^4}}{s(z_m)} \, \frac{A \pi}{z_m^2} \ .
\label{eq:Epoint}
\ee
The subsequent motion is sensitive to the value of this initial energy
of the $q\bar q$ pair.
The static potential between a heavy quark
and antiquark in $\Nfour$ SYM rises linearly at short distance
before switching to a Coulombic form at a cross-over distance
set by the inverse quark mass \cite{myers}.
At non-zero temperature, the potential remains approximately
Coulombic until a distance of order the inverse temperature, where
(at $\Nc=\infty$) there is an abrupt transition to
to a constant limiting value [which is twice the static thermal
rest energy $\Mrest (T)$] \cite{brandhuber, reyt}.%
\footnote
    {
    This corresponds to a change in the lowest energy string
    configuration from one in which a string connects the two quarks,
    to one in which a string runs straight down from each
    quark to the horizon.
    }
If the energy of the $q\bar q$ pair is sufficiently low, then the
attractive force between the separating quarks will be strong enough
to cause their trajectories to turn around, and the resulting motion
will resemble an oscillator.
If the energy is sufficiently high, then the quark trajectories will not have
turning points and the motion will resemble a highly overdamped oscillator.

For our point-like initial conditions,
the exact energy threshold for non-oscillatory motion
need not equal $2\Mrest(T)$ precisely, because of
the possibility of exciting internal string degrees of freedom.
Numerically, however, there does indeed appear to be a divergence
in the period at
$E \gtrsim 2 \Mrest(T)$ for these point-like initial conditions.

\begin{FIGURE}[t]
{
\centerline{
    \psfig{figure=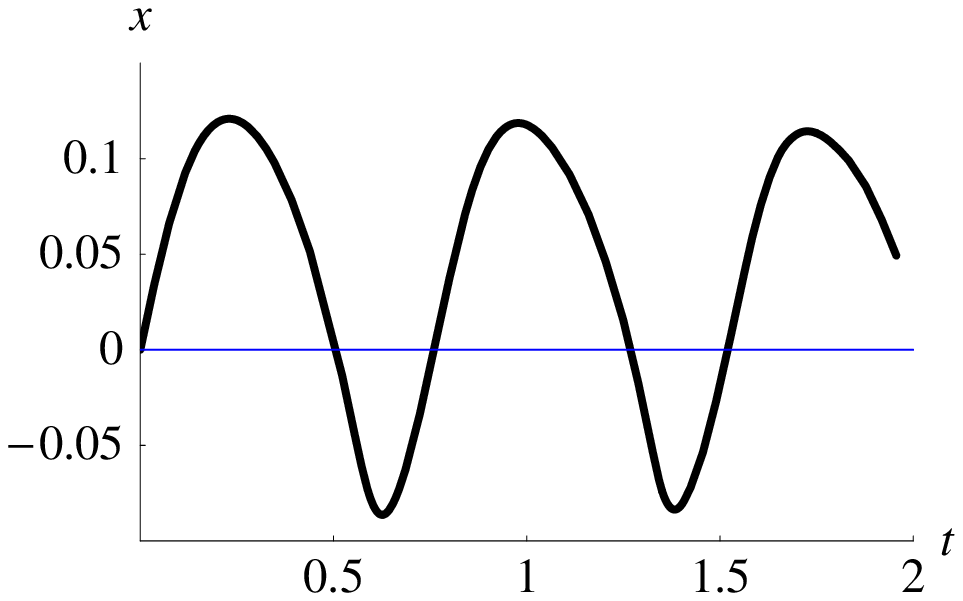,width=2.5in} \quad
    }
\vspace*{-1em}
\caption
    {
    String endpoint as a function of time for 
    an oscillatory solution with $z_m = 0.50$, $E/\Mrest = 1.2$, and pointlike
    initial conditions.  To observe many oscillations, this integration
    tolerated a larger numerical error 
    ($10^{-3}$ instead of $10^{-4}$ typical of the other plots in this section).
     The time dependent damping is comparable to the numerical error.
    }
\label{fig:oscillate}
}
\end{FIGURE}

Numerically we do find oscillating solutions similar to the
flat space expanding and contracting semi-circle (\ref{modRegge})
when $E$ is well below $2\Mrest(T)$.
An example is illustrated in Figure~\ref{fig:oscillate}.
As $E$ approaches $2\Mrest(T)$ from below, the period of oscillation
becomes longer and longer.

\begin{FIGURE}[t]
{
\centerline{
   \raisebox{3cm}{(a)}
    \psfig{figure=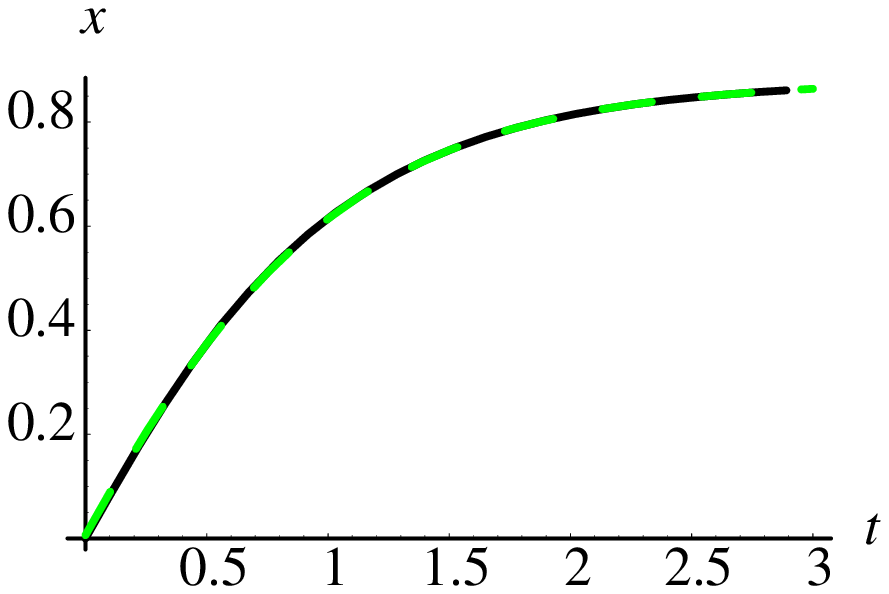,width=2.5in} \quad
   \raisebox{3cm}{(b)}
    \psfig{figure=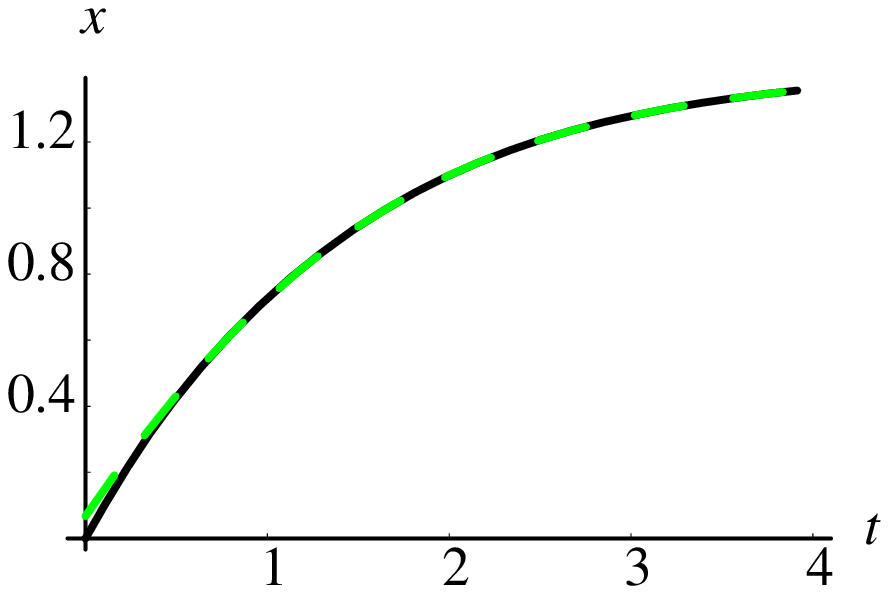,width=2.5in}}
\centerline{
   \raisebox{3cm}{(c)}
    \psfig{figure=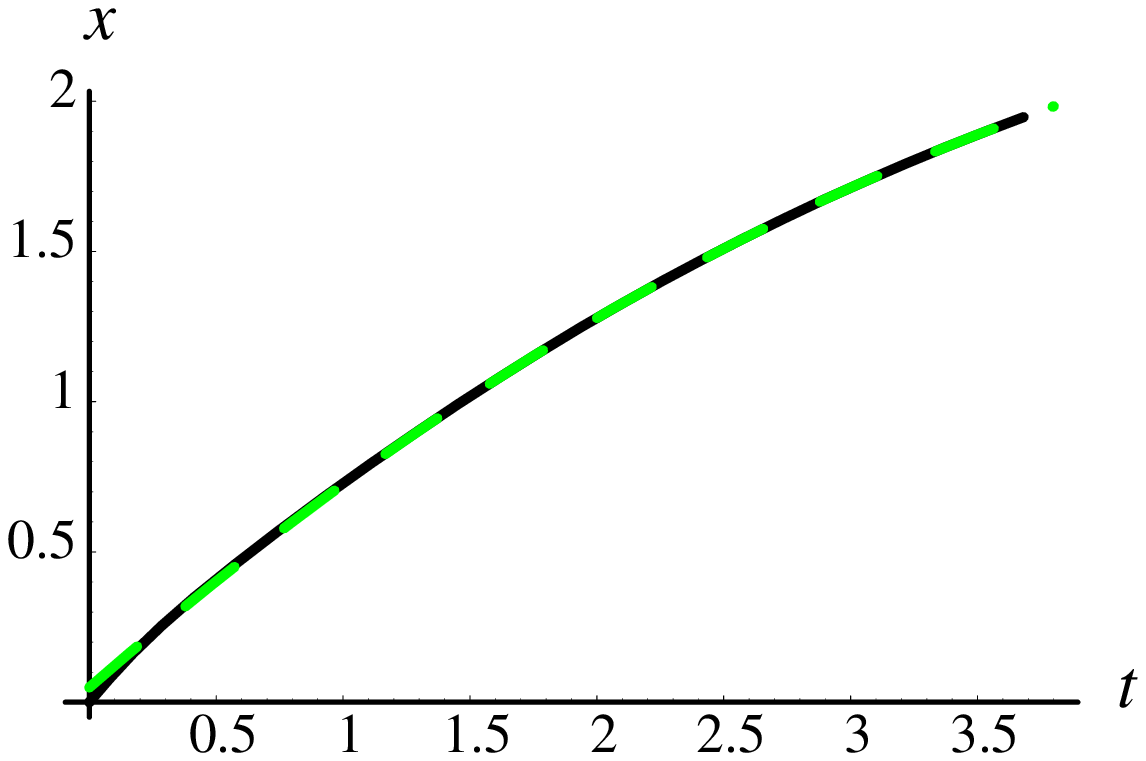,width=2.5in} \quad
   \raisebox{3cm}{(d)}
    \psfig{figure=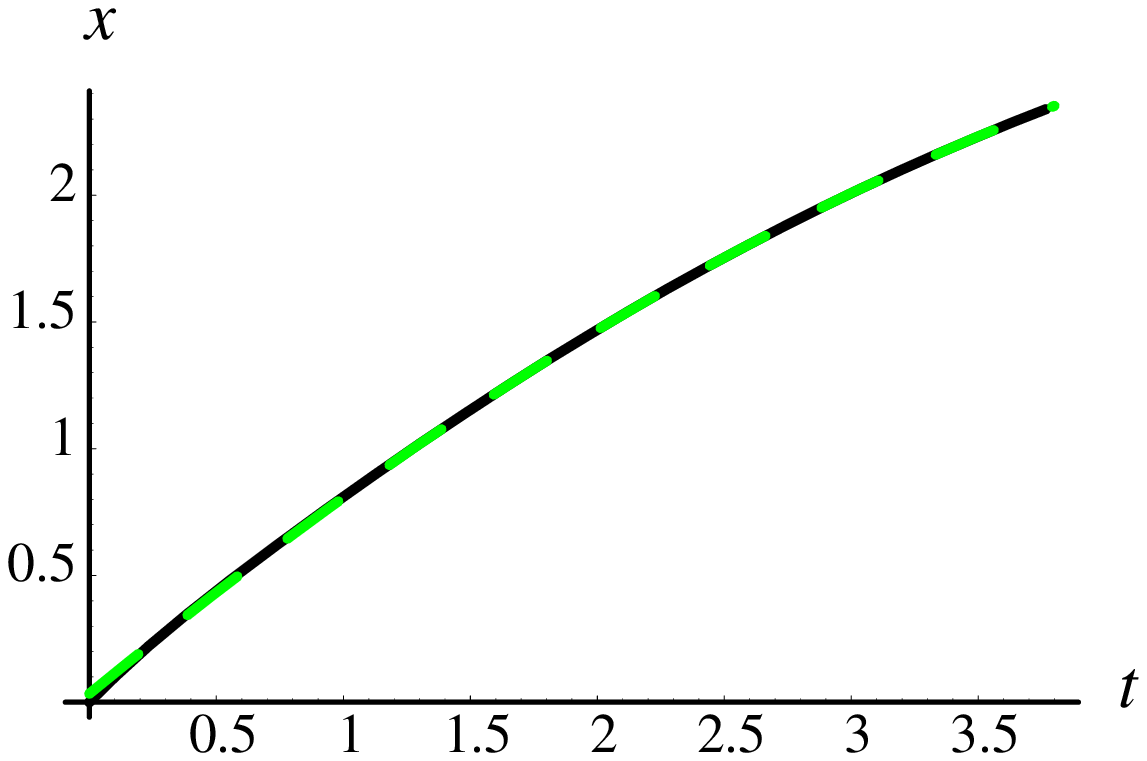,width=2.5in}}
\vspace*{-1em}
\caption{Plots of the position of the endpoint of the string versus time for
several different D7-brane positions $z_m$.
The string endpoint is plotted in black while the best-fit curve,
as described in the text, is a green dashed line.
The two curves are nearly identical.
Parameters for the different solutions are:
(a) $z_m=0.75$, $E/\Mrest=14$,
(b) $z_m=0.50$, $E/ \Mrest=6.1$,
(c) $z_m=0.25$, $E/\Mrest=4.4$, and
(d) $z_m=0.25$, $E/\Mrest=5.9$.
Units where $u_h \equiv \pi T = 1$ are used.
}
\label{seriesoffits}
}
\end{FIGURE}

To extract information about the viscous damping of a single quark,
we want to create a quark-antiquark pair with an energy greatly
exceeding the binding energy,
\be
    E \gg 2\Mrest(T) \,,
\label{Acond}
\ee
so as to
minimize the interaction between the quark and antiquark.
Numerical solutions satisfying this large energy condition
do show the expected non-oscillatory behavior.
Several examples are shown in Figure~\ref{seriesoffits}.
Solutions (a) and (b) used a
stretching factor $s=(1{-}z)$, amplitude $A=0.25$,
and flavor brane positions $z_m=0.75$ and 0.5 respectively.
The masses and energies were $m/\Delta m(T) = 1.28$,
$\Mrest(T)/m = 0.26$, and $E/\Mrest(T) = 14$ for (a) and
$m/\Delta m(T) = 1.98$,
$\Mrest(T)/m = 0.51$, and $E/\Mrest(T) = 6.1$ for (b).
For (c),
a stretching factor $s=(1{-}z)^{3/2}$, amplitude $A=0.17$,
and flavor brane position $z_m = 0.25$ were used.  For this
run,
$m/\Delta m(T) = 4.0$,
$\Mrest(T)/m = 0.75$, and $E/\Mrest(T) = 4.4$.
In the last run (d), a stretching factor of $s=(1{-}z)^2$, an amplitude $A=0.20$,
and a flavor brane position of $z_m=0.25$ were used, 
corresponding to  
$E/\Mrest(T) = 5.9$.
Numerical error limits how far we are able to integrate in time.
For initial conditions corresponding to lighter or less energetic quarks,
one sees a large decrease in velocity and a clear approach to
an asymptotically constant position.
But for higher energies or more massive quarks
(which experience less damping)
numerical error prevents us from following the quark to the
non-relativistic regime.

Given a quark-antiquark pair with large energy,
we model each quark independently as a particle
experiencing a damping force
\be
\frac{dp}{dt} = -\mu\,  p \,.
\label{eq:pdot}
\ee
If momentum is proportional to velocity, as with non-relativistic motion,
then this equation integrates to
\be
    v(t) = v_0 \, e^{-\mu t} \,, \qquad
    x(t) = x_\infty - \frac {v_0}{\mu} \, e^{-\mu t} \,.
\label{eq:nr}
\ee
But if the relation between velocity and momentum has a
relativistic form, $ p \propto v/{\sqrt{1-v^2}}$,
then equation (\ref {eq:pdot}) is equivalent to
\be
\frac{dv}{dt} = -\mu\,  v \, (1-v^2) \,,
\label{eq:vdot}
\ee
which integrates to
\begin{subequations}
\begin{eqnarray}
    v(t) &=& v_0 \bigm/ \sqrt{v_0^2 + (1-v_0^2) \, e^{2 \mu t}} \, ,
\\
\noalign{\hbox{and}}
    x(t) &=& x_\infty - \frac{1}{2\mu} \ln
    \left[
    \frac{1+v(t)}{1-v(t)}
    \right] .
\label{fitting}
\end{eqnarray}
\label{eq:fitrel}
\end{subequations}
To test the validity of this description and extract information
on the damping rate $\mu$
we fit the numerical results for the position of the string
endpoint to Eq.~(\ref{eq:fitrel}),
treating $\mu$, $\ln(1-v_0^2)$ and $x_\infty$ as free parameters.%
\footnote
{
Given a numerical integration up to a worldsheet time $\tau_{max}$, we fit
the endpoint of the string to the assumed form for $x(t)$ only for
$\tau_{max}/2 < \tau < \tau_{max}$,
choosing ten equally spaced points in this region.
We limit the data used in the fit to the latter half of the available
time interval in order to minimize the effects of the interaction
between the quarks, which is largest when the quarks are close together.
}
As shown in Figure~\ref{seriesoffits},
when the large energy condition (\ref{Acond}) holds
the resulting fit, using the form (\ref {eq:fitrel}),
is quite good for all $\tau$,
although we do see some minor deviations at small $\tau$.
In contrast, fits of these large energy solutions to the simple
exponential form (\ref{eq:nr}) are somewhat worse.

The fact that fits using Eq.~(\ref{eq:fitrel})
are so good supports the presumed relativistic relation
between $p$ and $v$ together with a friction coefficient
$\mu$ that is independent of $p$.
We are following the evolution of
the quark over typically a rather large momentum range as can be seen
from Figure~\ref{seriesoffits}.
The results for the extracted values of $\mu$ are nearly independent
of the energy so long as $E > 2\Mrest(T)$.
Changing the energy by 50\% or more
typically only results in a few percent change in the value of $\mu$.
For example, for a D7-brane at $z_m=0.5$,
as we increase $E/\Mrest$ from 4.8 to 7.2,
$\mu$ changes from 0.79 to 0.80.
For the case $z_m=0.25$, changing $E/\Mrest$ from 4.4 to 5.9
changes $\mu$ from 0.325 to 0.326.

\begin{TABLE}[t]
{
\centerline{
\begin{tabular}{ccccc}
&&  \multicolumn{2}{c}{$\quad\mu/ \pi T$} \\
\raise 8pt \hbox{\smash{$\displaystyle\frac {m}{\Delta m(T)}$}}
& \raise 7pt \hbox{$z_m$} & numeric & QNM  \\
\hline
5.00 & 0.20 & 0.250 & 0.25 \\
4.00 & 0.25 & 0.325 & 0.32 \\
3.12 & 0.32 & 0.44 & 0.44 \\
2.49 & 0.4 & 0.59 & 0.59 \\
1.98 & 0.5 &  0.80 & 0.80 \\
1.64 & 0.6 & 1.02 & 1.04 \\
1.28 & 0.75 & 1.40 & 1.42 \\
\end{tabular}
}
\caption{
The friction coefficient $\mu$ for various values of quark mass $m$.
The quasinormal mode (QNM) results were calculated using the linear
analysis of Section~\protect\ref{sec:qnm},
while the numeric results come from the
full time dependent numerics discussed in this Section.
}
\label{table:visc}
}
\end{TABLE}

Table~\ref{table:visc} compares the values of $\mu$ from the quasinormal mode
calculation of Section~\ref{sec:qnm} to the best fits of our numerical
integrations.
The numbers are astonishingly close, giving us confidence
in the linear analysis.
The results begin to differ by a few percent for small mass quarks.
It is not a-priori clear what causes this discrepancy.
The extraction of the energy loss rate may be affected by the interaction
with the other quark,
but that should produce a correction of the opposite sign.
There may be small thermal corrections to the
relation between momentum and velocity, or residual
momentum dependence in the damping rate.
Conceivably, there could be nonlinear effects that are absent in the quasinormal
mode analysis but which reappear in this full dynamical simulation
and are relevant even at (accessibly) late times.
Or this small discrepancy may
reflect residual errors in our numerical integration.

\section{Discussion}\label{sec:discussion}

Let us close with a discussion of the validity of our approximations. The
classical treatment of the string is justified as long as the string is
much longer than a string length; quantum fluctuations will be suppressed
by powers of $\ell_s/R$ where $R$ denotes the characteristic length of
the string. All the single quark solutions we considered had strings
with length of order $L$, the AdS curvature radius, or larger.
As one lowers the quark mass toward the critical value
$m_{\rm c} \approx 0.92 \, \Delta m(T)$, the
D7-brane approaches but does not quite reach the horizon.
The shortest string one can get (with $m \ge m_{\rm c}$)
has a length of about $\ucrit L$.
Since $L = \lambda^{\frac{1}{4}} \, \ell_s$, we
see that for sufficiently large $\lambda$, quantum fluctuations of the
string are always suppressed.
Hence, a classical treatment of the string dynamics is valid
for sufficiently large $\lambda$ as long as the quark mass exceeds
the critical value $m_{\rm c}$.

For applications to QCD, however, one may be interested in
large but not asymptotically huge values of the 't Hooft coupling,
perhaps $\lambda \approx 20$ (corresponding to $\alphas \approx 0.5$).
In this regime, the condition $R\gg \ell_s$ can become non-trivial and
for masses near the critical value $m_{\rm c}$
(which is about 2.2 $T$ for $\lambda \approx 20$)
quantum fluctuations of the string will be important.

Another phenomenon 
that has not appeared in our discussion up to now is Brownian motion.
Any dissipative thermal system must also have fluctuations,
as shown by the fluctuation-dissipation theorem.
In particular, a quark (of finite mass) initially at rest in the plasma
should not be able to remain at rest.
It will undergo Brownian motion and diffuse away from
its starting point.
Over a time $t$ it will travel a distance $\Delta x \sim \sqrt{ D t}$.
The diffusion constant $D$ is directly related to the viscous drag,
\begin{equation}
    D = \frac {T}{\mu M}
    \,.
\label {eq:D}
\end{equation}
This physics is missing in our classical treatment of string dynamics.
A motionless string stretching from the D7-brane to the horizon
is a solution to the equations of motion,
and there is no obvious reason for the string endpoint to move at all.
This straight motionless string clearly represents a quark
at rest in the plasma.

The reason we do not see Brownian motion is due to the
non-uniform nature of the large $\lambda$ and large time limits.
For a quark initially moving with some $O(1)$ velocity $v$,
stochastic Brownian motion will be unimportant until sufficiently late times.
To see this explicitly, one can
use the Langevin equation
\be
\dot {\vec p} = -\mu \, \vec p + \vec \xi(t)
\label{eq:langevin}
\ee
to model the behavior of the quark.
The required assumptions underlying this description are discussed below.
Here $\vec\xi(t)$ is stochastic white noise, with variance
\begin{equation}
\langle \xi_i(t) \xi_j(t') \rangle = C \, \delta_{ij} \, \delta(t{-}t') \,.
\label{eq:noise}
\end{equation}
Calculating the mean square momentum at time $t$ gives
\be
    \langle p_i(t) p_j(t) \rangle
    =
    \langle p_i(t) \rangle\langle p_j(t) \rangle
    + \frac{C}{2\mu} \, \delta_{ij} \, (1-e^{-2\mu t}) \,.
\label{P2sol}
\ee
with $\langle \vec p(t) \rangle = \vec p(0) \, e^{-\mu t}$.
At equilibrium, by equipartition,
the kinetic energy $p^2/(2 \Mkin)$ of the quark must be $\coeff 32 T$.
(Note that the condition $m \ge m_{\rm c}$ automatically implies that
$\Mkin \gg T$, so a non-relativistic form of kinetic energy is
appropriate for quarks in equilibrium.)
Requiring that the large $t$ limit of $\langle \vec p(t)^2 \rangle$
equal $3T \Mkin$ shows that
the strength of the noise is determined by the viscous drag,
\begin{equation}
    C = 2T \mu \Mkin(T)
    = 2\pi T^2 \Delta m(T)
    = \pi \sqrt\lambda \, T^3
    \,.
\label{eq:C}
\end{equation}
Integrating the Langevin equation (\ref{eq:langevin}) again to find
the quark's position, assuming non-relativistic motion and vanishing
initial velocity,
and computing the mean square displacement gives
\begin{equation}
    \langle \Delta x_i(t) \, \Delta x_j(t) \rangle
    =
    \frac{C \, t }{(\mu \Mkin)^2}  \; \delta_{ij}
    \big[1 + \mathcal O\big(1/(\mu t)\big)\big] \,.
\label{eq:diffuse}
\end{equation}
This must equal the classic result
$
    \langle \Delta x_i(t) \, \Delta x_j(t) \rangle
    = 2D \, t \, \delta_{ij}
$
for the variance of the probability distribution
$
    P(\Delta\vec x,t) =
    (4\pi D t)^{-3/2} \,
    e^{-\Delta\vec x^2/(4Dt)}
$
generated by
a diffusion equation with diffusion constant $D$.
Combining Eqs.~(\ref{eq:C}) and (\ref{eq:diffuse}) gives
the stated value (\ref{eq:D}) for the diffusion constant.

For diffusive effects to be negligible, we require the second term of
(\ref{P2sol}) to be small compared
to the first term, giving an upper bound on the time
over which a deterministic treatment of quark motion is valid,
\be
t < t_B \equiv \frac{1}{2\mu} \ln \left( 1 + \frac{2 K.E.}{3T} \right) ,
\ee
where $K.E. = \half \Mkin \, v_0^2$
is the kinetic energy of the quark at time zero.
At the same time, for Eq.~(\ref{eq:langevin}) to adequately model the
energy loss of the quark,
quantum uncertainty in its kinetic energy must be negligible compared
to the change in the kinetic energy which we wish to resolve.
This imposes a lower limit on the time during which our classical
description is valid,
namely
$1/t \ll \mu t \, (K.E.)$, or
\begin{equation}
    t \gg t_Q \equiv \left[ {\mu \times (K.E.)} \right]^{-1/2} \,.
\end{equation}

Our analysis requires that $t_Q \ll t \ll t_B$.
Using $\mu \sim \Delta m(T) T / \Mkin$,
this condition may be rewritten as
\begin{equation}
    \sqrt {\frac {\Delta m}{\Mkin}}
    \ll
    \sqrt{\frac {K.E.}{T}} \>
    \ln \left( 1 + \frac{2 K.E.}{3T} \right) .
\end{equation}
This inequality is most stringent for the lightest (accessible) quarks
with $m$ near $m_{\rm c}$ and $\Mkin \approx \half \Delta m(T)$.
Hence the required large separation between the quantum and diffusive
time scales is valid as long as the
initial kinetic energy of the quark is large compared to the temperature,
\begin{equation}
    K.E. \equiv \half \Mkin \, v_0^2 \gg T \,.
\end{equation}
For a quark mass near the critical limit $m_{\rm c}$,
this is equivalent to the requirement that the initial velocity satisfy
\begin{equation}
    v_0 \gg \lambda^{-1/4} \,.
\end{equation}
A point to be emphasized is that the classical treatment of quark
dynamics underlying all results in Table~\ref{table:visc} 
is valid provided $\lambda$ is sufficiently large.

The result (\ref{P2sol}) for momentum fluctuations
also allows one to relate the
rate of change of the mean square transverse momentum
to the noise, and hence to the viscous drag,
\begin{equation}
    \frac d{dt}
    \left\langle \vec p_\perp(t)^2 \right\rangle \Big|_{t=0}
    =
    2C
    =
    4T \mu \Mkin(T)
    = 2\pi \sqrt\lambda \, T^3 \,.
\label{eq:transverseD}
\end{equation}
This characterizes the diffusion in transverse momentum of a quark
propagating through the plasma.
However, the simple connection (\ref{eq:transverseD}) between transverse
momentum diffusion and viscous drag
relies on the validity of the Langevin equation
(\ref {eq:langevin})
with isotropic white noise (\ref{eq:noise})
for modeling the stochastic force fluctuations acting on the quark.
If the stochastic force fluctuations have significant
momentum dependence, or non-Gaussian correlations,
then this simple description will not be adequate.
One can argue that the simple Langevin description is valid
for non-relativistic motion, $v \ll 1$.
Whether it remains valid for arbitrary momentum is not completely clear.
At weak coupling, both the viscous drag and the noise variance
acquire significant velocity dependence for $\mathcal O(1)$
values of rapidity \cite{MooreTeaney}.
Since we find no velocity dependence in the friction coefficient $\mu$
at strong coupling,
it seems plausible that there will also be negligible velocity
dependence in the variance of the force fluctuations,
even for relativistic motion.
However, this has not been directly verified.

\vskip 0.1in
\noindent {\bf Note Added:}

Several related papers \cite{MIT,SolanaTeaney,Gubser}
have very recently appeared which overlap with portions of our analysis.
The quark diffusion constant found in Ref.~\cite{SolanaTeaney}
agrees with our value (\ref{diffusionresult}).
The result of Ref.~\cite{MIT} for the jet quenching parameter $\hat q$
does not agree with our result (\ref{jetquenching}).
However, these authors are addressing a different physical
question involving radiative energy loss of a lightlike projectile.
Interestingly, their result has the same parametric dependence
as our result (\ref{eq:transverseD}),
but with a coefficient which is 15\% smaller.

\section*{Acknowledgments}
We would like to thank A.~O'Bannon, S.~Minwalla, and M.~Strassler
 for useful discussions.
This work was supported in part by the U.S. Department
    of Energy under Grant No.~DE-FG02-96ER40956
    and by the National Science Foundation under Grant No. PHY99-07949.

\appendix

\section{Other solutions}
\label{app:other}

In this appendix we briefly discuss other stationary
string solutions in the $AdS_5$-black hole background.
As we showed in Section~\ref{sec:analytic},
the ansatz $x(u,t)=x(u) + v t$
reduces the string equation of motion to the ordinary differential equation
(\ref{ourdiffeq}) with a first integral (\ref {oursol}).
The numerator in the expression (\ref{oursol}) vanishes at
$u_c = (1-v^2)^{-1/d} u_h$,
while (for $d=4$) the denominator vanishes at
$u_0 = (u_h^4 + C^2 v^2)^{1/4}$.
The previously discussed single quark solution requires $u_c=u_0$,
so that $x'$ is non-vanishing and non-singular everywhere between
the horizon and the boundary.
But if these two radii do not coincide,
one can still find solutions with $-g$ everywhere positive on the
worldsheet.
If $u_0>u_c$, then the physical solution lives entirely
in the $u>u_0$ part of space, depicted on the left in
Figure~\ref{twilson}.
This configuration corresponds to an infinitely heavy external
quark/antiquark pair at finite temperature, moving at a common velocity $v$.
For a static quark/antiquark pair, it was found in
Ref.~\cite{brandhuber,reyt} that the solution only exists
for a bounded range of quark/antiquark separations $l$.
At zero velocity the largest separation
$l_{max}$ is achieved when $u_0$ approaches $u_h$.
At non-zero velocity, this solution ceases to exist beyond $u_0=u_c$.

Alternatively, if $u_0<u_c$ then the solution lives entirely in the
region between $u_0$ and the horizon, as depicted
on the right in Figure~\ref{twilson}.
One might think 
that this string solution could represent some coherent gluonic excitation.
But since the momentum
is outgoing on one end of the string, and ingoing on the
other, we believe this solution is unphysical.

\begin{FIGURE}
{
   \centerline{\psfig{figure=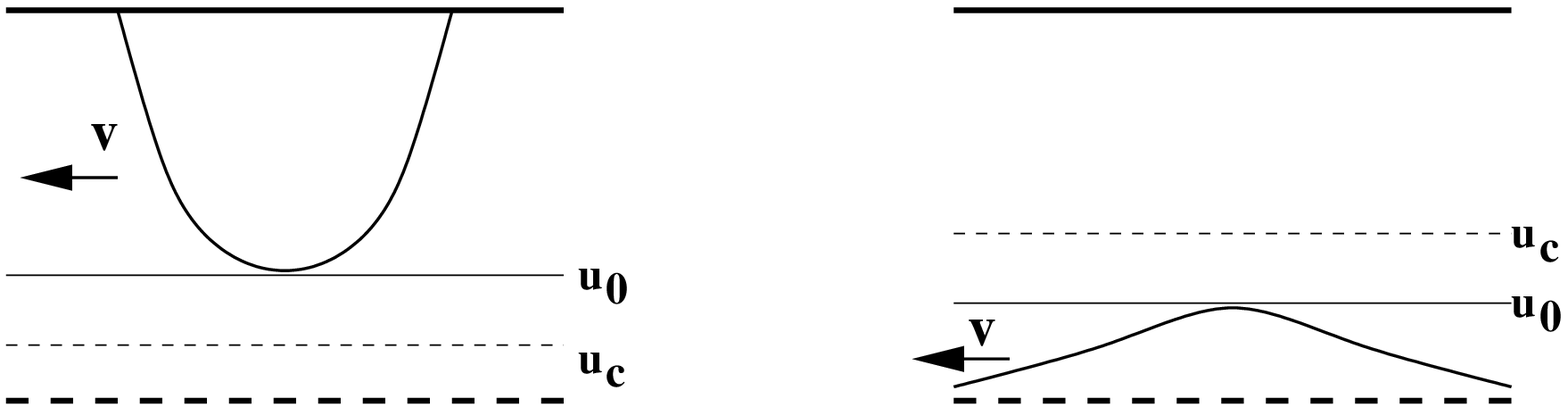,width=5.0in}}
    \caption{Left:
    Semiclassical solutions corresponding to a quark/antiquark
    pair with a fixed spatial separation moving through the finite
    temperature medium at constant speed.
    Right: A stationary solution in which the string moves at constant
    velocity outside the horizon.
    One end of the string satisfies physical (outgoing) boundary conditions
    at the horizon, but the other end does not.
    Hence, this solution is unphysical.}
\label{twilson}
}
\end{FIGURE}

\begin{FIGURE}[t]
{
   \centerline{\psfig{figure=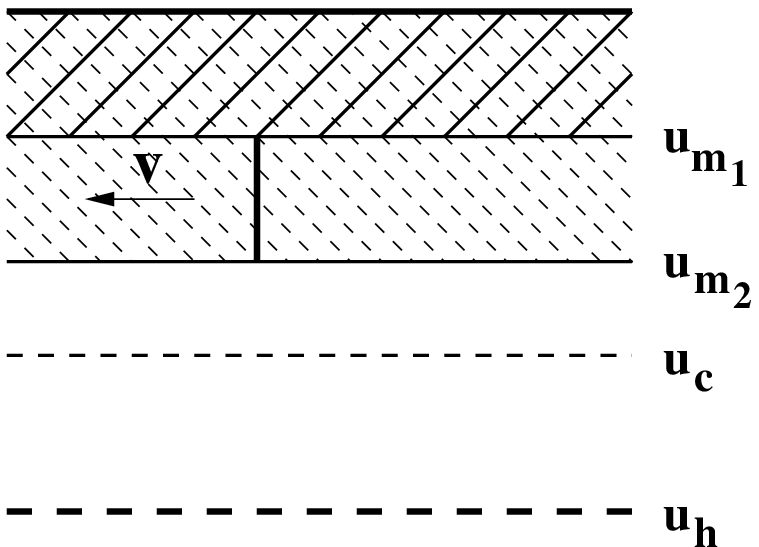,width=2.6in}}
    \caption{A straight, moving string solution
    corresponding to a light-heavy meson (in a multi-flavor theory)
    moving through the thermal medium at constant velocity.
    }
\label{fig:light-heavy}
}
\end{FIGURE}

Finally,
yet another very simple solution is a straight string, $x = vt$, moving at
constant velocity.
We argued in Section \ref{sec:single} that such a string,
stretching from $u_m$ down to the horizon and moving at any non-zero
velocity, is not physical.
However, if one considers a theory with two flavors of quarks
with different masses, so that the gravitational description has two
D7-branes at differing radial positions $u_{m_1}$ and $u_{m_2}$,
then one can regard the portion of this trivial solution
lying between $u_{m_1}$ and $u_{m_2}$ as describing a moving
light-heavy meson.
Figure~\ref{fig:light-heavy} depicts this configuration.
The solution is physical provided $u_{m_1}$ and $u_{m_2}$ are both
greater than $u_c$.
This condition shows that, for given quark masses, there is a maximum
velocity with which such a color-singlet meson can move through the
medium without experiencing any drag
(at leading order in $1/N_c$ and $\lambda\to\infty$),
\begin{equation}
    v^2 < 1 - (u_h/u_{m_<})^d \,,
\end{equation}
where $u_{m_<}$ is the lesser of $u_{m_1}$ and $u_{m_2}$.

\newpage

\section{Quasinormal modes in $d=2$}
\label{app:qnm}

In $d=2$, the metric function $h(u) = u^2 - u_h^2$
and the resulting linear equation (\ref{linearizedeomt})
can be solved analytically. The most general solution is
\beq x(y,t) = e^{-\mu t} f(y) =
e^{- \mu t} \left ( c_P \, P_1^{\wn}(y)/y + c_Q \, Q_1^{\wn}(y)/y
\right ) ,
\eeq
where we have introduced the dimensionless quantities $\wn=\mu/u_h$
and $y=u/u_h$,
and $P^\gamma_1$ and $Q^\gamma_1$ are associated Legendre functions.
(We follow the conventions in Gradshteyn and Ryzhik \cite{GR}.)
The horizon is at $y=1$, and the boundary is out at $y \rightarrow \infty$.
In order to study the behavior of $P^\gamma_1$ and $Q^\gamma_1$ at large $y$, and for
$y$ close to 1, it is convenient to use the relation to
hypergeometric functions. The associated Legendre functions
$P^\gamma_1$ and $Q^\gamma_1$ are uniquely defined
for $1<y<2$ and $y>1$, respectively, as
\begin{eqnarray}
P^{\wn}_1(y) &=& \frac{1}{\Gamma(1{-}\wn)} \left ( \frac{y+1}{y-1}
\right )^{\wn/2}
{}_2 F_1 \left(-1,2,1{-}\wn ;  \half {-} \coeff y{2} \right) \,,
\\
Q^{\wn}_1(y)
&=& \coeff 13 { e^{i \wn \pi} \, \Gamma(2{+}\wn)}
\left(1-y^{-2} \right)^{\wn/2}
y^{-2} \;
{}_2 F_1
\left(\coeff \wn 2 + \coeff 3 2, \coeff \wn 2+1,\coeff 5 2 ; y^{-2} \right)  .
\end{eqnarray}
These expressions may be used to evaluate $P^\gamma_1(y)$ for $y$ close to 1
and $Q^\gamma_1(y)$ at large $y$.
To evaluate $P^\gamma_1(y)$ at large $y$ and $Q^\gamma_1(y)$ close to 1
one can use hypergeometric
identities to find (see also Ref.~\cite{GR}):
\begin{eqnarray}
\nonumber
P^{\wn}_1(y) &=& \frac{1}{3 \Gamma(-1{-}\wn)} \left(1-y^{-2}\right)^{\wn/2}\,
y^{-2} \;
{}_2 F_1
\left(\coeff \wn 2{+}\coeff 3 2, \coeff\wn 2 {+} \half,\coeff 5 2 ; y^{-2}\right)
\\&& {} +
\frac{1}{\Gamma(2{-}\wn)} \left(1-y^{-2} \right)^{\wn/2} \, y \;
 {}_2 F_1 \left(\coeff \wn 2,\coeff\wn 2{-}\half,-\half ; y^{-2} \right) ,
\\
\nonumber
Q^{\wn}_1(y) &=& \half {e^{i \wn \pi}}
\biggl [
    \Gamma(\wn) \left ( \frac{y+1}{y-1} \right )^{\wn/2}
    {}_2 F_1 \left(-1,2,1{-}\wn ; \half{-}\coeff y{2}\right)
\\&& \qquad {} +
    \frac{\Gamma(-\wn) \Gamma(2{+}\wn)}{\Gamma(2{-}\wn)}
     \left (\frac{y-1}{y+1} \right )^{\wn/2}
     {}_2 F_1 \left(-1,2,1{+}\wn ; \half{-}\coeff y{2} \right)
\biggr ]  \ .
\end{eqnarray}
Close to the horizon we find that $Q^\gamma_1$ is a linear combination of
a solution that goes as $(y{-}1)^{\wn/2}$ and one that goes as
$(y{-}1)^{-\wn/2}$.
According to Eq.~(\ref{nha}),
when combined with $e^{- \mu t}$ time dependence
the former is in-going while the latter is outgoing.
On the other hand,
$P^\gamma_1$ only has a $(y{-}1)^{-\wn/2}$ term and hence is
the purely outgoing solution. So to find quasinormal modes we can focus
on the $P^\gamma_1$ solution only,%
\footnote{To check the results
we also looked at $e^{\mu t}$ time dependence instead of $e^{-\mu t}$.
In that case $(y{-}1)^{\wn/2}$ is
the physical near horizon behavior. This implies a particular linear
combination of $P^\gamma_1$ and $Q^\gamma_1$.
The final answer turns out to be the same.
}
and set $c_Q=0$.

Imposing Neumann
boundary conditions at a flavor brane, the
quasinormal modes are given by solutions to
\beq f'(y) \big|_{y=y_m} = \partial_y (P_1^{\wn}(y)/y) \big|_{y=y_m} =0.
\eeq
For large $y_m$ we can expand the hypergeometric functions to obtain
\begin{equation}
 f(y) = \frac{1}{\Gamma(2{-}\wn)} \left [ 1- \half {\wn^2} y^{-2} 
+{\cal O} \left({\wn^4}y^{-4} \right) \right ]
+
\frac{y^{-3}}{3 \Gamma(-1 {-}\wn) }
\left [
1 + \coeff 1{10}(3 + 9 \wn + \wn^2)\, y^{-2} + {\cal O} \left(y^{-4} \right)
\right] .
 \end{equation}
Assuming that $\wn \sim 1/y_m$, we have kept all terms up to order $y_m^{-7}$.
This assumption will be justified presently.
From this asymptotic expansion, it follows that
\beq
\label{full}  f'(y_m) = 
\frac{y_m^{-3}}{\Gamma(-1{-}\wn)}
\left[
\frac{\wn}{1{-}\wn^2} \left( 1+\coeff{1}{2}\, y_m^{-2} \right)
- y_m^{-1} - \left( \half {+} \coeff{3 \wn}{2} \right) y_m^{-3} +
{\cal O} \left(y_m^{-5} \right)
\right] .
\ee
Solving $f'(y_m)=0$ for $y_m$ yields the asymptotic expression
$
y_m = \frac{1}{\wn} - \frac{\wn}{2} + {\cal O} (\wn^2)
$,
which justifies {\it a posteriori} our assumption
about the scaling behavior of $\wn$.
Inverting this expression for $\wn$ yields
\be
\wn = \frac{1}{y_m} - \frac{1}{2\,y_m^3} + {\cal O} \left( y_m^{-4}\right) \,.
\ee
The corresponding quasinormal mode wavefunction $f(y)$ is plotted
in Figure~\ref{legendre} for $y_m=10$.

\begin{FIGURE}[t]
{
   \centerline{\psfig{figure=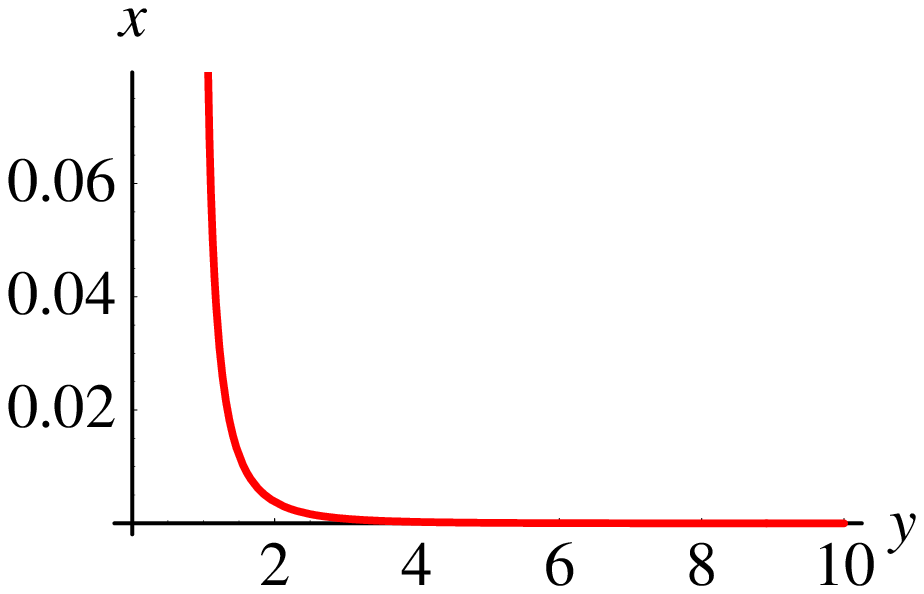,width=2.4in}}
    \caption{Quasinormal mode wavefunction $f(y)$ for a mass
corresponding to $y_m=10$.}
    \vspace*{-2em}
\label{legendre}
 }
 \end{FIGURE}


\section{Numerical error}
\label{app:error}

To perform the numerical integration in Section~\ref{sec:numerics},
we used the {\tt NDSolve} routine provided by Mathematica \cite{Mathematica} and checked for
numerical error in a variety of ways.
We used the spatial error estimate provided by {\tt NDSolve},
we checked the constraint equations,
and we compared the numerical integration results for different grid spacings.

The routine
{\tt NDSolve} produces a warning if its internal spatial error estimate
exceeds a threshold.  None of the numerical results we present here 
generated such warnings.
This spatial error estimate is formed by considering the final step in
the numerical propagation.
One additional time step is made both with the original grid and with a
coarser grid with half the number of grid points.
Using the Richardson extrapolation formula,
Mathematica produces a warning if the difference
$
\frac{|y_2 - y_1|}{2^p - 1} > 10
$,
where $y_2$ and $y_1$ are the resulting values of the function for the
two different choices of grid and $p$ is the order of the
discretization routine that converts derivatives into differences,
the default value of which is $p=4$.
The norm $|y_2{-}y_1|$ involves a scaled sum over the difference between
$y_2$ and $y_1$ which the documentation of {\tt NDSolve} does not
describe in detail.

We made our own rough estimates of the numerical error by numerically
integrating each example twice with two different grids.  If the first
grid was specified to have a minimum number of points $n$,
the second grid would have a minimum of $2n$ points.
The (absolute) difference between the two integrations was kept under about
$10^{-4}$.

The constraint equations (\ref{constraints}), if satisfied by
the initial conditions, should  remain satisfied at all later times.
But unless special steps are taken when converting derivatives into finite
differences,
the differencing scheme will no longer preserve the constraint equations
exactly.  The extent to which the constraint equations are satisfied at
later times is thus an indirect measure of the accumulated numerical error.
In the numerical results presented here, we checked the constraints at a
handful of points and found that they differed from zero only by
about $10^{-6}$.

\newpage

\bibliography{gkp}
\bibliographystyle{JHEP}

\end{document}